\documentclass[11pt,journal]{IEEEtran}

\usepackage[mathcal]{euscript}
\usepackage{epsfig}
\usepackage{amsmath,amstext,amsfonts,amssymb}
\usepackage{epsfig,float}
\usepackage{graphicx}
\usepackage{subfigure}
\usepackage{url}
\usepackage{shadow,color,pifont,times,rotate}
\onecolumn

\newcommand{\ul}{\underline}

\newcommand{\bs}{\boldsymbol}
\newcommand{\mc}{\mathcal}

\newcommand{\ds}{\displaystyle}

\newtheorem{theorem}{Theorem}[section]

\newtheorem{proposition}[theorem]{Proposition}

\begin{document}

\title{\Large {\bf Gaussian Broadcast Channels with an Orthogonal and Bidirectional Cooperation Link \footnote{Some results
concerning the AF protocol have been presented at the $8^{th}$ IEEE
Signal Processing Advances in Wireless Communications workshop
(SPAWC), June 2007, Helsinki, Finland.}}}


\author{E.~V. Belmega, B. Djeumou and S. Lasaulce
\begin{center}
Laboratoire des Signaux et Syst\`{e}mes\\
CNRS -- Sup\'{e}lec -- Paris 11\\
3 rue Joliot-Curie -- 91192 Gif-sur-Yvette Cedex -- France\\
{belmega / djeumou / lasaulce}@lss.supelec.fr\\
\today
\end{center}}

\maketitle

\begin{abstract}
This paper considers a system where one transmitter broadcasts a
single common message to two receivers linked by a bidirectional
cooperation channel, which is assumed to be orthogonal to the
downlink channel. Assuming a simplified setup where, in particular,
scalar relaying protocols are used and channel coding is not
exploited, we want to provide elements of response to several
questions of practical interest. Here are the main underlying
issues: 1. The way of recombining the signals at the receivers; 2.
The optimal number of cooperation rounds; 3. The way of cooperating
(symmetrically or asymmetrically; which receiver should start
cooperating in the latter case); 4. The influence of spectral
resources. These issues are considered by studying the performance
of the assumed system through analytical results when they are
derivable and through simulation results. For the particular choices we made,
the results sometimes do not coincide with those available for the
discrete counterpart of the studied channel.

\end{abstract}


\IEEEpeerreviewmaketitle

\section{Introduction}
\label{intro}

In the conventional broadcast channel (BC) introduced by
\cite{cover-1972}, one transmitter
 sends independent messages to several receivers. The channel under
  investigation in
this paper differs from the original BC for at least two reasons.
First, the receivers can cooperate in
 order to enhance the overall system performance. Second, the
  users want to decode the
same message. We will refer to this situation as the cooperative
broadcast channel (CBC) with a single common message. For the sake of
simplicity a 2-user CBC will be assumed. Note that the considered
channel is also different from the original relay channel (RC)
defined in \cite{cover-it-1979}, because each terminal not only
acts as a relay but also as a receiver, which means that
ultimately, the information message has to be decoded by both
terminals. Additionally the cooperation channel between the two
receivers is assumed to be bidirectional (versus unidirectional
for the RC) and orthogonal to the downlink (DL) channels. Although
their sub-optimality, orthogonal channels are often assumed for
practical reasons (e.g. it is difficult/impossible to implement
relay-receivers that receive and transmit at the same
 time in the same frequency band).

To the author's knowledge the most significant
contributions\footnote{For example the authors note that
\cite{liang-wcsit-2005}\cite{liang-it-2007} also addressed the CBC
but did not focus on the common message case.} concerning the
situation under investigation are
\cite{draper-allerton-2003}\cite{azarian-it-2005}\cite{ng-itw-2006}\cite{lasaulce-icassp-2006}\cite{khalili-allerton-2006}\cite{dabora-it-2006}.
The \emph{discrete} broadcast channel with a bidirectional
conference link\footnote{The exact original definition of a
conference link is given in \cite{willems-it-1983}. It essentially
consists of a noiseless channel with a finite capacity.} and a
single common message was originally studied by Draper et al. in
\cite{draper-allerton-2003}. The authors proposed a way of decoding
the message in multiple rounds and applied their scheme to the
binary erasure channel. The corresponding coding-decoding scheme is
based on the use of auxiliary variables while a certain form of
channel comparability\footnote{Commenting on this concept is out of
the scope of this paper. For more information see
\cite{korner-hungary-1975}\cite{korner-it-1977}\cite{elgamal-it-1979}.
Example: The channel $p(y_{1}|x)$ is said to be less noisy than
$p(y_2|x)$ if for any auxiliary random variable $U$, $I(U;Y_1) \geq
I(U;Y_2)$. The main point here is that the achievable rates of
\cite{draper-allerton-2003} are not derived in the general case but
assuming certain Markov chains.} is assumed through these variables.
This channel has also been analyzed by \cite{dabora-it-2006} where
the authors essentially proposed achievable rates based on the use
of estimate-and-forward (EF) at both receivers and two-round
cooperation schemes. The Gaussian counterpart of this channel has
been studied in \cite{lasaulce-icassp-2006}. Showing the optimality
of decode-and-forward for an unidirectional cooperation, the authors
evaluated the exact loss due to the channel orthogonalization. For
the bidirectional case, the proposed achievable rate is based on a
combination of EF and decode-and-forward (DF) and shown to always
outperform the pure EF-based solution (always for the 2-round
decoding). Independently \cite{ng-itw-2006} exploited a similar
approach to analyze the Gaussian relay channel with a bidirectional
cooperation. The fading case has been partially treated in
\cite{azarian-it-2005}. The diversity-multiplexing trade-off,
achieved by using a ``dynamic'' version of decode-and-forward, is
derived for the unidirectional cooperation case.

While the authors of
\cite{azarian-it-2005}\cite{lasaulce-icassp-2006}\cite{dabora-it-2006}
addressed situations where only one or two cooperation exchanges (or
decoding rounds) are allowed, this paper focuses on the case where
the number of exchanges is arbitrary. For the erasure channels,
\cite{draper-allerton-2003} and \cite{khalili-allerton-2006} have
shown that the higher the number of exchanges the better the
performance in terms of information rate. However the discrete
channel analysis (including erasure channels) does not take into
consideration the spectral resources aspect. As it will be seen,
this point is in fact crucial and accounting for it can lead to
markedly different conclusions from
\cite{draper-allerton-2003}\cite{khalili-allerton-2006} concerning
the optimum number of cooperation exchanges. Additionally,
\cite{draper-allerton-2003} and \cite{khalili-allerton-2006} only
considered the information rate as a performance criterion whereas
other criteria of interest can also be considered. Although assuming
special cases of relaying protocols, this paper aims precisely at
taking into account these two aspects for providing some insights to
the following issues:
\begin{enumerate}
\item The way of recombining the signals at the receivers. Indeed,
the receiver can combine the cooperation signal with either its
downlink signal or the combiner output from the previous iteration.
Also, the choice of the combining scheme (which depends on the
assumed relaying protocol) will also be discussed.

\item The optimal number of the cooperation rounds. In contrast with the
discrete case this number will be shown to be less than or equal to
$2$ if the cooperation protocols are properly chosen.

\item The way of cooperating. The choice between symmetric and
asymmetric can be made based on a simple discussion but it will also be
illustrated by numerical results. Simulations will also indicate the
relative importance of the order in which the receivers start to
cooperate.

\item The influence of the spectral resources on the three mentioned
issues will be assessed. Two different assumptions are made:
($\mc{H}_1$) The total system bandwidth is fixed; ($\mc{H}_2$) Only
the downlink channel bandwidth is fixed.
\end{enumerate}

In order to provide elements of response to these questions we will
use a simplified approach. After presenting the used system model
(sec. \ref{sec:system-model}), we will evaluate the exact equivalent
signal-to-noise ratio (SNR) in the output of the maximum ratio
combiner (MRC) for each user (sec. \ref{sec:af}) in the case where a
scalar, memoryless and zero-delay amplify-and-forward protocol (AF)
is assumed at both receivers. This will be done for several
cooperation strategies. In order to assess the influence of the
relaying protocol on the aforementioned issues we will also evaluate
the system performance when DF is used at the relay. In this case,
 a more sophisticated combiner (namely a maximum
likelihood detector -- MLD), which is provided in Sec. \ref{sec:df},
has to be used at the receivers. Based on the choice of different
system performance criteria (sec. \ref{sec:perf-crit}), numerical and
simulation analyses will be conducted (sec. \ref{sec:experimental}).
Concluding remarks and possible extensions of the present work will
be provided in Sec. \ref{sec:concl}.

\begin{figure}[!h]
\centering
\includegraphics[width=0.5\columnwidth]{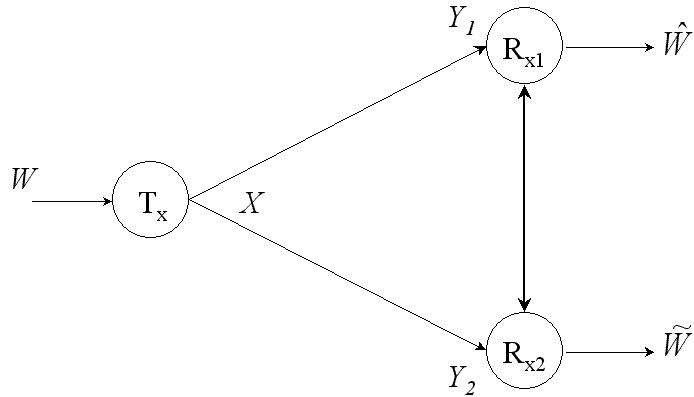}
\caption{\footnotesize{The cooperative broadcast channel with a
single common message ($W$) and an orthogonal and bidirectional
cooperation channel.}} \label{fig-channel}
\end{figure}
%

\section{System model}
\label{sec:system-model}

As mentioned in the previous section, the system under investigation
(see Fig. \ref{fig-channel}) comprises one transmitter (source) and
two receivers (destinations). The transmitted signal is denoted by
$X$ and subject to a power constraint: $\mathbb{E}|X^2| \leq P$. Its
bandwidth is denoted by $B_{DL}$. For the sake of simplicity, $X$ will be
assumed to be a scalar quantity e.g. a Gaussian input or a
quadrature amplitude modulation (QAM) symbol. Assuming an additive
white Gaussian noise (AWGN) model for the different links of the
system, the baseband downlink signals write:
\begin{equation}
\left\{
\begin{array}{ccl}
Y_1 & = & X + Z_1 \\
Y_2 & = & X + Z_2
\end{array}
\right.
\end{equation}
where for all $j \in \{1,2 \}$, $Z_j \sim \mathbb{C} \mc{N}(0, n_j
B_{DL})$, $n_j$ is the noise power spectral density for receiver
$j$, and $I(Z_1;Z_2) = 0$. We will assume that orthogonality between
the downlink and cooperation channels is implemented by frequency
division (FD). The bandwidth allocated to the cooperation channel
between the two receivers is denoted by $B_{C}$. The cooperation
channel can be divided into several sub-channels, each of them
having a bandwidth equal to $\Delta B$. The two receivers cooperate
by applying the same relaying strategy namely either the AF or DF
protocol. Using the AF protocol imposes the condition $\Delta B =
B_{DL}$ whereas $\Delta B $ and $B_{DL}$ can be chosen independently
(or possibly through a compatibility
 constraint between the source and relay data rates) when DF is used for relaying . Regarding
 the spectral resources aspect, two different scenarios will be
 considered. In the first scenario we assume that $B_{DL} + B_{C} =
 const$ (\emph{Assumption} $\mc{H}_1$). This corresponds to the situation where the total system bandwidth is
fixed, which is generally assumed to fairly compare two systems
\emph{before} implementation. In the second scenario we assume that
$ B_{DL} = const$ (\emph{Assumption} $\mc{H}_2$), which does not
lead to fair comparisons in terms of bandwidth since the cooperation
channel bandwidth can be chosen arbitrarily. The attention of the reader
is drawn to the fact that, although unfair, this scenario still
makes sense in the real life. For instance, consider the case where
one wants to assess the benefits of cooperation by coupling two
\emph{existing} communication systems such as a DVB (digital video
broadcasting) system and a cellular system. As modifying the DVB
system downlink signal bandwidth would be a difficult/an impossible
task, the second assumption, which amounts to extending the
available bandwidth is more appropriate for comparing a DVB system
with its terrestrial cooperation-based counterpart.

At last, we will assume scalar and zero-delay relaying. In real
situations, this can be implemented for instance by re-synchronizing
the downlink and cooperation signals at the receivers. The main
advantage for assuming scalar protocols is that the additional
complexity caused by the cooperation is low, it does not imply large
decoding delays and some analytical results can be derived. As in
\cite{dabora-it-2006}, two main ways to cooperate are distinguished
in this paper: the symmetric and asymmetric cooperation. The main
distinction between these cooperation types is that for the
symmetric cooperation the receivers exchange their cooperation
signal simultaneously, while in the asymmetric cooperation the
exchanges are done sequentially \textit{i.e.} one receiver sends a
cooperation signal at a given time. In the case where each receiver
amplifies and forwards its received downlink signal the symmetric
cooperation can be seen as a special case of the asymmetric
cooperation.


\section{The case of amplify-and-forward}
\label{sec:af}

\subsection{Selected combining scheme}

Let us consider the first cooperation round for the symmetric cooperation. Each receiver (\emph{e.g.} receiver 1) amplifies and
forwards his received downlink signal ($ Y_1$ for receiver 1) to his partner (receiver 2). This is done simultaneously. Then
each receiver (say receiver 2) has to combine its downlink signal with the cooperation signal received from his partner
($Y_{12}^{(1)} = a_{12}^{(1)} Y_1 + Z_{12}^{(1)}$). To combine these signals we chose the MRC. The motivation for this choice
is threefold.  First, one of the features of the MRC is that it is simple. The MRC has also two properties. By definition it
maximizes the equivalent SNR at its output. As shown in Appendix \ref{sec:proof1} it also maintains the mutual information
constant. The data processing theorem indicates that the MI between $X$ and the MRC output has
    to be less than or equal to the MI between $X$ and its (vector) input. It turns out that for the
    choice of weights maximizing the equivalent SNR, there is no
    loss of MI. At last, an additional motivation for the MRC is
    that it can be proved that using a more advanced combiner such as the MMSE will bring
    nothing more by taking into account the structure and statistics of
    the different signals. Now consider the second iteration of the cooperation procedure. Each
receiver has at least two choices in terms of cooperation signals to be sent: it can continue to send its original downlink
signal (\emph{Strategy} $\mc{S}_2$) or it can send the MRC output from the previous iteration (\emph{Strategy} $\mc{S}_1$). The
first ($\mc{S}_1$) strategy is the counterpart of the strategy presented in \cite{dabora-it-2006} for the discrete CBC.
Normally this strategy is intended to be better than the second one ($\mc{S}_2$) since the receiver can ``de-noise'' or remove
some wrong information bits from the estimated data flow. Here, in our simplified setup (channel decoding is not exploited),
the goal is to prove the intuition that sending to your partner what you received from him cannot improve the performance,
which ultimately means that the second strategy is better than the first one.

\subsection{Received signals}
\label{af-asym-sym}

 Consider the case of the symmetric cooperation. To
denote the signals of interest for a given cooperation round or
iteration $i$, with $i \in \{1, ..., K_s\}$, we will use the
following notations:
\begin{equation}
\label{eq:received-af} \left\{
\begin{array}{ccll}
Y_{I}^{(i)} & = & \alpha_{I}^{(i)} X + Z_{I}^{(i)} \\
Y_{II}^{(i)} & = & \alpha_{II}^{(i)} X + Z_{II}^{(i)}  \\
Y_{12}^{(i)} & = & a_{12}^{(i)} Y_{I}^{(j)} + Z_{12}^{(i)}  \\
Y_{21}^{(i)} & = & a_{21}^{(i)} Y_{II}^{(j)} + Z_{21}^{(i)}
\end{array}
\right.
\end{equation}
where $Y_{I}^{(i)}$ (resp. $Y_{II}^{(i)}$) corresponds to the MRC
output at iteration $i$ and receiver 1 (resp. receiver 2),
$a_{12}^{(i)}$, $a_{21}^{(i)}$ are the scalar AF protocol
amplification gains, which are determined by the total cooperation
powers available: $P_{12}$
 at receiver 1 and $P_{21}$ at receiver 2. At last, $Y_I^{(0)} =
 Y_1$, $Y_{II}^{(0)} =
 Y_2$, $j=i-1$ for the strategy $\mc{S}_1$ and $j=0$ for the strategy $\mc{S}_2$.
For the asymmetric cooperation, we will keep the same
notations for the signals of interest as in the symmetric case. However,
in contrast with the symmetric cooperation, combining operations take place
at receiver 2 for odd indices $i$ only, and at receiver 1 for even
indices $i$ only (under the assumption that receiver 1 starts
relaying).

Whereas the notations are identical for the asymmetric and symmetric
cooperation, the bandwidth of the cooperation channel is defined
differently. If one denotes by $K_s$ the number of pairs of
cooperation exchanges in the case of symmetric cooperation we have
\begin{equation}
\Delta B = \left|
\begin{array}{ccccccc}
\ds{\frac{B}{2 K_s + 1}} & \mathrm{when} & B_{DL} + B_{C} &= &const.
& \triangleq & B  \\
B & \mathrm{when} &   B_{DL} &= &const. & \triangleq & B,
\end{array}
\right.
\end{equation}
and if one denotes by $K_a$ the number of cooperation exchanges in
the case of asymmetric cooperation we have
\begin{equation}
 \Delta B = \left|
\begin{array}{ccccccc}
\ds{\frac{B}{K_a + 1}} & \mathrm{when} & B_{DL} + B_{C} &= &const.
&\triangleq & B \\
B & \mathrm{when} &   B_{DL} &= &const.&\triangleq & B.
\end{array}
\right.
\end{equation}

%
%
%

\subsection{Equivalent SNR analysis}
\label{af-snr}

The purpose of this section is to evaluate analytically the
equivalent SNR at the MRC output after an arbitrary number of
cooperation rounds for the two mentioned strategies. This allows us
 not only to compare them in terms of the SNR, but also to use this knowledge to evaluate
 other performance criteria presented in Sec. \ref{sec:perf-crit}.

\subsubsection{The case of the strategy $\mc{S}_1$}

In this case, it turns out that it is not possible, in general, to
express the equivalent SNR as a function of the sole channel
parameters ($P, P_{12}, n_1, ...$). In fact the equivalent SNR has
to be determined recursively. The purpose of Theorem \ref{theo-snr}
(see Appendix \ref{sec:proof2} ) is precisely to provide this
relationship, both for the asymmetric and the symmetric cooperation types. Before
providing this theorem and the two underlying propositions, we need
to mention and detail one important point regarding the interest in
these results. First, let us consider the case where the system
bandwidth is fixed. Imposing $\Delta B = \frac{B}{K+1}$ (with $K =
K_a$ or $K=2K_s$ depending on the context) allows us to perform fair
comparisons in terms of spectral resources whatever the value for
$K$. However, the cases $K=0$, $K=1$ and $K=2$ do never correspond
to fair comparisons in terms of power since they respectively
correspond to $(P, P_{12} = 0, P_{21} = 0)$, $(P, P_{12}, P_{21} =
0)$ and $(P, P_{12}, P_{21})$. For $K \geq 2$ the comparisons are
spectrally fair because the total cooperation powers are kept fixed.


\begin{theorem} \label{theo-snr}
\emph{(General expression for the equivalent SNRs). Assume that $n_1
< n_2 $ and receiver 2 performs the MRC task in the first place if
asymmetric cooperation is considered. For iteration $i \in
\{1,...,K\}$ the corresponding weights are denoted by $w_{2}^{(i)}$
(weighting the MRC output at iteration $i-1$) and $w_{12}^{(i)}$
(weighting the cooperation signal). For receiver 1 the weights are
denoted by $w_{1}^{(i)}$, $w_{21}^{(i)}$. Denote by $Y_{I}^{(i)} =
\alpha_{I}^{(i)} X + Z_{I}^{(i)}$ (resp. $Y_{II}^{(i)} =
\alpha_{II}^{(i)} X + Z_{II}^{(i)}$) the signal at MRC output for
receiver 1 (resp. receiver 2) and iteration $i$, with $Z_{I}^{(i)}
\sim \mc{N}(0, N_{I}^{(i)})$ (resp. $Z_{II}^{(i)} \sim \mc{N}(0,
N_{II}^{(i)})$). Let $\rho_{I}^{(i)}$ (resp. $\rho_{II}^{(i)}$) be
the signal-to-noise ratio associated with the signal $Y_{I}^{(i)}$
(resp. $Y_{II}^{(i)}$). The SNRs $\rho_{I}^{(i)} \triangleq \ds{
\frac{\mc{S}_{I}^{(i)}}{\mc{T}_{I}^{(i)}} }$ and $\rho_{II}^{(i)}
\triangleq \ds{ \frac{\mc{S}_{II}^{(i)}}{\mc{T}_{II}^{(i)}} }$ can
be determined recursively as follows:
\begin{equation}
\begin{array}{ccl}
\mc{S}_{II}^{(i)}  &=& \ds{\alpha_{I}^{(i-1)}\alpha_{II}^{(i-1)}
\left(e^{(i-1)}+e^{(i-1),*}\right)\rho_{I}^{(i-1)}\rho_{II}^{(i-1)}\rho_{12}}
 \ds{-\left(\alpha_{I}^{(i-1)}\alpha_{II}^{(i-1)}\right)^2P
 \left[\rho_{II}^{(i-1)}\left(1+\rho_{I}^{(i-1)}\right) \right.} \\
 && \ds{+ \left.
 \rho_{12}\left(\rho_{I}^{(i-1)}+\rho_{II}^{(i-1)}\right)\right]}\\
\mc{T}_{II}^{(i)} & = & \ds{
\frac{e^{(i-1)}e^{(i-1),*}}{P}\rho_{I}^{(i-1)}\rho_{II}^{(i-1)}\rho_{12}
} \ds{- \left(\alpha_{I}^{(i-1)}\alpha_{II}^{(i-1)}\right)^2P\left(1
+ \rho_{12}\right)} - \ds{\left(\alpha_{I}^{(i-1)}\right)^2
 N_{II}^{(i-1)}  \rho_{I}^{(i-1)}\rho_{II}^{(i-1)}}
\end{array}
\end{equation}
 where $\rho_{12} = \ds{\frac{P_{12}}{n_{12} \Delta B}}$, $k$
is a constant depending on the cooperation scheme (asymmetric or
symmetric), $(.)^*$ denotes the conjugate, $e^{(0)} = 0$,
$N_{I}^{(0)} = N_1$, $N_{II}^{(0)} = N_2$, $\rho_{I}^{(0)} =P/N_1$,
$\rho_{II}^{(0)} =P/N_2$, $\alpha_{I}^{(0)} = \alpha_{II}^{(0)} = 1
$. The amplification gains are defined by: $a_{12}^{(i)}
=\sqrt{\frac{P_{12}^{(i)}}{( \alpha_{I}^{(i-1)}
    )^2P+N_{I}^{(i-1)}} }$, $a_{21}^{(i)} =\sqrt{\frac{P_{21}^{(i)}}{(
     \alpha_{II}^{(i-1)}
    )^2P+N_{II}^{(i-1)}} }$  and $P_{12}^{(i)}$, $P_{21}^{(i)}$ are the
    available cooperation powers per subchannel.
For the SNR $\rho_{I}^{(i)}$ do the following changes for the
indices: $I \leftrightarrow II$ and $1 \leftrightarrow 2$.}
\end{theorem}

The expressions of the signals coefficients $\alpha_{I}^{(i)}$,
$\alpha_{II}^{(i)}$, the cooperation powers per subchannel
$P_{12}^{(i)}$ , $P_{21}^{(i)}$ and the equivalent noise powers
$N_{I}^{(i)}$, $N_{II}^{(i)}$  depend on the cooperation type.
Expressing these quantities is the purpose of the following two
propositions.


\begin{proposition}
\label{prop-symm} \emph{(MRC weights for the symmetric cooperation).
For the symmetric cooperation the MRC weights can be shown to be:
\begin{equation}
\label{eq:weights1} \left\{
\begin{array}{ccl}
 w_{12}^{(i)} &= &  a_{12}^{(i)}\alpha_{I}^{(i-1)}N_{II}^{(i-1)}-  a_{12}^{(i)}
 \alpha_{II}^{(i-1)} e^{(i-1)}
 \\
 w_{2}^{(i)} & = & \left[(a_{12}^{(i)})^2N_{I}^{(i-1)}+N_{12}^{(i)} \right]
 \alpha_{II}^{(i-1)} - (a_{12}^{(i)})^2 \alpha_{I}^{(i-1)} e^{(i-1),*} \\
 w_{21}^{(i)} &= &  a_{21}^{(i)}\alpha_{II}^{(i-1)}N_{I}^{(i-1)}-  a_{21}^{(i)}
 \alpha_{I}^{(i-1)} e^{(i-1),*}
 \\
 w_{1}^{(i)} & = & \left[(a_{21}^{(i)})^2N_{II}^{(i-1)}+N_{21}^{(i)} \right]
 \alpha_{I}^{(i-1)} - (a_{21}^{(i)})^2 \alpha_{II}^{(i-1)} e^{(i-1)}
\end{array}
\right.
\end{equation}
where}
\begin{itemize}
    \item \emph{$e^{(i-1)} \triangleq E \left[ Z_{I}^{(i-1)} Z_{II}^{(i-1),*}  \right]$
     with
\begin{equation}
\begin{array}{ccl}
e^{(i)} & = &   w_{12}^{(i)} a_{12}^{(i)} w_{1}^{(i)}
N_{I}^{(i-1)} + w_{21}^{(i)} a_{21}^{(i)} w_{2}^{(i)} N_{II}^{(i-1)}\\
& &   + \left[ w_{1}^{(i)}   w_{2}^{(i)} + w_{12}^{(i)} a_{12}^{(i)}
w_{21}^{(i)} a_{21}^{(i)} \right] e^{(i-1)},
\end{array}
\end{equation}}
    \item \emph{for all $i \in \{1,...,K_s \}$ the useful signal coefficients are
given by
\begin{equation}
\left\{
\begin{array}{ccl}
\alpha_{I}^{(i)} &= &w_{21}^{(i)}a_{21}^{(i)}\alpha_{II}^{(i-1)} +
w_{1}^{(i)} \alpha_I^{(i-1)} \\
\alpha_{II}^{(i)} &= &w_{12}^{(i)}a_{12}^{(i)}\alpha_{I}^{(i-1)} +
w_{2}^{(i)} \alpha_{II}^{(i-1)},
\end{array}
\right. \end{equation}}
    \item \emph{the cooperation powers per subchannel are for
    all $i \in \{1,...,K_s\}$ given by
\begin{equation}
\left\{
\begin{array}{ccc}
 P_{12}^{(i)} & = & \ds{\frac{P_{12}}{K_s} }\\
 P_{21}^{(i)} & = & \ds{\frac{P_{21}}{K_s} },
\end{array}
\right. \end{equation}}
    \item \emph{the equivalent noise powers $N_{I}^{(i)}$,
    $N_{II}^{(i)}$ are determined through
\begin{equation}
\left\{
\begin{array}{ccc}
Z_{I}^{(i)} &=& w_{21}^{(i)}a_{21}^{(i)}Z_{II}^{(i-1)} +
w_{21}^{(i)}Z_{21}^{(i)}+w_{1}^{(i)}Z_{I}^{(i-1)} \\
Z_{II}^{(i)} &=& w_{12}^{(i)}a_{12}^{(i)}Z_{I}^{(i-1)} +
w_{12}^{(i)}Z_{12}^{(i)}+w_{2}^{(i)}Z_{II}^{(i-1)}.
\end{array}
 \right.
\end{equation}}
    \item \emph{for all $i \in \{1,...,K_s \}$: $N_{12}^{(i)}  = n_{12}
    \Delta B
    $ and $N_{21}^{(i)}  = n_{21} \Delta B $}

    \item \emph{the constant $k$ of Theorem \ref{theo-snr} equals $\ds{\frac{2 K_s +
    1}{K_s}}$.}

\end{itemize}
\end{proposition}

\begin{proposition} \label{prop-asymm}
\emph{(MRC weights for the asymmetric cooperation). For the
asymmetric cooperation the MRC weights can be shown to coincide with
that of Proposition \ref{prop-symm} where}
\begin{itemize}
    \item \emph{$e^{(i-1)} \triangleq E \left[ Z_{I}^{(i-1)} Z_{II}^{(i-1),*}
    \right]$ with
\begin{equation}
e^{(i)}
 =
 \left|
\begin{array}{cc}
w_{1}^{(i)} e^{(i-1)} + w_{21}^{(i)} a_{(21)}^{(i)} N_{II}^{(i-1)}
 & \ for \ i \ even \\
w_{2}^{(i)} e^{(i-1)} + w_{12}^{(i)} a_{(12)}^{(i)} N_{I}^{(i-1)} &
\ for \ i \ odd,
\end{array}
 \right.
\end{equation}}
    \item \emph{the useful signal coefficients are given by
\begin{equation}
\alpha_{I}^{(i)} = \left|
\begin{array}{cc}
w_{21}^{(i)}a_{21}^{(i)}\alpha_{II}^{(i-1)}+w_{1}^{(i)}
\alpha_{I}^{(i-1)} & \ for \ i \ even \\
\alpha_{I}^{(i-1)} & \ for \ i \ odd,
\end{array}
 \right.
\end{equation} }
\emph{\begin{equation} \alpha_{II}^{(i)} = \left|
\begin{array}{cc}
\alpha_{II}^{(i-1)} & \ for \ i \  even \\
w_{12}^{(i)}a_{12}^{(i)}\alpha_{I}^{(i-1)}+w_{2}^{(i)}\alpha_{II}^{(i-1)}
& \ for \ i \ odd,
\end{array}
\right.
\end{equation} }

    \item \emph{the cooperation powers per subchannel are for
    all $i \in \{1,...,K_a\}$
\begin{equation}
P_{12}^{(i)} = \left|
\begin{array}{cc}
\ds{\frac{2 P_{12}}{K_a} }&
for \ K_a \ even \\
\ds{\frac{2 P_{12}}{K_a+1}} & \ for \ K_a \ odd,
\end{array}
 \right.
\end{equation}} \emph{\begin{equation} P_{21}^{(i)} = \left|
\begin{array}{cc}
\ds{\frac{2 P_{21}}{K_a}} &
\ for \ K_a \ even \\
\ds{\frac{2 P_{21}}{K_a-1}} & \ for \ K_a \ odd, \ K_a \geq 3,
\end{array}
 \right.
\end{equation}}
    \item \emph{the equivalent noise powers $N_{I}^{(i)}$,
    $N_{II}^{(i)}$ are determined through
\begin{equation}
Z_{I}^{(i)} = \left|
\begin{array}{cc}
w_{21}^{(i)}a_{21}^{(i)}Z_{II}^{(i-1)} +
w_{21}^{(i)}Z_{21}^{(i)}+w_{1}^{(i)}Z_{I}^{(i-1)}
&  \ i \ even \\
Z_{I}^{(i-1)} & \ i \  odd,
\end{array}
 \right.
\end{equation}}

\emph{
\begin{equation}
Z_{II}^{(i)} = \left|
\begin{array}{cc}
Z_{II}^{(i-1)}
&  \ i \ even \\
w_{12}^{(i)}a_{12}^{(i)}Z_{I}^{(i-1)} +
w_{12}^{(i)}Z_{12}^{(i)}+w_{2}^{(i)}Z_{II}^{(i-1)}
 & \ i \  odd,
\end{array}
 \right.
\end{equation}}
    \item \emph{for all $i \in \{1,...,K_a\}$: $N_{12}^{(i)}  = n_{12}
    \Delta B
    $ and $N_{21}^{(i)}  = n_{21} \Delta B $,}
\end{itemize}

    \item  \emph{the constant $k$ of Theorem \ref{theo-snr} equals 1.}

\end{proposition}
The theorem and propositions provided here are proved in Appendix
\ref{sec:proof2},\ref{sec:proof3},\ref{sec:proof4}.

\subsubsection{The case of the strategy $\mc{S}_2$}

As the strategy $\mc{S}_2$ consists in always sending to the other
receiver the downlink signal, it can be easily checked that the
performance of the symmetric case with a number of pairs of
cooperation rounds equal to $K_s$ is the same as the asymmetric
case with $2 K_s$ cooperation rounds. As the symmetric case is
easier to expose and the derivations in both cases are similar, we
restrict our attention to the symmetric case here. The received
signal are particularly simple to express in the case of strategy
$\mc{S}_2$:
\begin{equation} \left\{
\begin{array}{ccll}
Y_{I}^{(i)} & = & \ds{\left(w_1 + \sum_{i=1}^{K_s}
w_{21}^{(i)}a_{21}^{(K_S)}\right)X + Z_1
+ \sum_{i=1}^{K_s} w_{21}^{(i)}Z_2 + Z_{21}^{(i)} }\\
Y_{II}^{(i)} & = & \ds{\left(w_2 +
\sum_{i=1}^{K_s}w_{12}^{(i)}a_{12}^{(K_s)}\right)X + Z_2 +
\sum_{i=1}^{K_s} w_{12}^{(i)}Z_1 + Z_{12}^{(i)}}
\end{array}
\right.
\end{equation}
where it can be checked that $w_1 = \frac{1}{N_1}$, $w_{21}^{(1)} =
\ldots = w_{21}^{(K_s)} =\frac{a_{21}^{(K_s)}}{K_s(a_{21}^{(K_s)})^2
N_2 + N_{21}}$, $w_2 =\frac{1}{N_2}$ and $w_{12}^{(1)} = \ldots =
w_{12}^{(K_s)} = \frac{a_{12}^{(2)}}{K_s(a_{12}^{(2)})^2N_1 +
N_{12}}$. We obtain that
\begin{equation}
\label{sN} \left\{
\begin{array}{cclclcl}
\rho_{I}^{(i)} & = & \left[\frac{1}{N_1} +
\frac{K_s(a_{21}^{(K_s)})^2}{K_s(a_{21}^{(2)})^2N_2 +
N_{21}}\right]P & = & \left[\frac{1}{N_1} +
\frac{(a_{21}^{(1)})^2}{(a_{21}^{(1)})^2N_2 +
N_{21}}\right]P & = & \rho_1 + \rho_{21}^{eff}\\
\rho_{II}^{(i)} & = & \left(\frac{1}{N_2} +
\frac{K_s(a_{12}^{(K_s)})^2}{K_s(a_{12}^{(2)})^2 N_1 +
N_{12}}\right)P & = & \left[\frac{1}{N_2} +
\frac{(a_{12}^{(1)})^2}{(a_{12}^{(1)})^2N_1 + N_{12}}\right]P & = &
\rho_2 + \rho_{12}^{eff}
\end{array}
\right.
\end{equation}
where the equalities at the right come from $a_{(12)}^{(K_s)} =
\frac{a_{12}^{(1)}}{\sqrt{K_s}}$ and $a_{(21)}^{(K_s)} =
\frac{a_{21}^{(1)}}{\sqrt{K_s}}$, with $a_{12}^{(1)} =
\sqrt{\frac{P_{12}}{P_1 + N_1}}$ and $a_{21}^{(1)} =
\sqrt{\frac{P_{21}}{P_2 + N_2}}$.

The main observation to be made here is that, if we consider the
case of the fixed downlink channel bandwidth (this case also implies
that $N_1$, $N_2$, $N_{12}$ and $N_{21}$ are independent of the
number of cooperation exchanges), the equivalent SNRs do not depend
on the cooperation round index for $i \geq 2$. Therefore the average
effect brought by the MRC is exactly compensated by the loss in
terms of cooperation power per exchange, the latter being translated
by the amplification gains $a_{12}^{(i)} =
\frac{a_{12}^{(i)}}{\sqrt{K_s}}$, $a_{21}^{(i)} =
\frac{a_{21}^{(i)}}{\sqrt{K_s}}$.

\subsubsection{Comparison of the two strategies}

The ideal result we would like to obtain is to determine the sign of
$\rho_{I, \mc{S}_2 }^{(i)} - \rho_{I, \mc{S}_1}^{(i)}$ for any
cooperation round index $i$. It turns out that this is not easy and
the underlying expressions become more and more complicated as $i$
increases. Therefore we chose to explicit the aforementioned
difference in a specific case but the reasoning can be applied to
other case of interest. For the asymmetric case (the most general one)
with $K_a =2$ and when the downlink bandwidth is constant, one can
show that the numerator of $\rho_{I, \mc{S}_2 }^{(i)} - \rho_{I,
\mc{S}_1}^{(i)}$ expresses as:

\begin{equation}
\begin{array}{ccl}
\mathrm{Num}\left(\rho_{I, \mc{S}_2 }^{(i)} - \rho_{I,
\mc{S}_1}^{(i)}\right) &=& P N_2P_{21}P_{12} \times  \\
& &
\left(2N_{21}N_{12}P^2+PN_{21}N_{12}N_2+2PN_1N_{21}N_{12}+PP_{21}N_{12}N_2+2PN_1P_{12}N_{21}+
\right.
 \\
& & \left. PP_{12}N_{21}N_2+N_1N_{21}N_{12}N_2+N_1P_{21}N_{12}N_2+N_1P_{12}N_{21}N_2\right) \\
& \geq & 0.\end{array}\label{eq:N-diff}
\end{equation}
This result shows that for two cooperation rounds, it is better for
the partner to send his downlink signal than the MRC output.
Simulation results will allow us to better quantify this difference
for any number of cooperation rounds.

\section{The case of decode-and-forward}
\label{sec:df}

\subsection{Differences between the AF and DF cases}


In Sec. \ref{sec:af} we assumed a scalar AF protocol for cooperation
between the two receivers. For the considered scenario we calculated the
equivalent SNR at the MRC output, after an arbitrary number of
cooperation exchanges. This calculation did not require any
assumption on the signals transmitted by the source and the relays. In
particular, a Gaussian signal could be assumed at the source and
relays and therefore the equivalent SNRs could be used to obtain an
achievable transmission rate for the considered system. In this
section we assume \emph{finite} modulations at the source and relays
(typically QAM modulations). Now the relay tries to recover the
source information messages and re-encodes and re-modulates them
into symbols to be sent to the destination. Ideally, these symbols
would be the source symbols. Therefore one can define, for each
relay, a discrete input discrete output channel between the source
and each relay output. The transition probabilities of each of these
channels are directly linked to the considered downlink channel SNR
and the error correction capacity of the decoder.

Assuming decode-and-forward type protocols at the relays implies
three main differences between the AF and DF cases.

\begin{enumerate}
    \item The MRC is the optimum combiner when AF is assumed for
    relaying. When a DF-type relaying protocol is assumed, some
    decoding noise is introduced by the relay, which is not
    compensated for by the MRC. As the simulation results of
    \cite{djeumou-isspit-2006} show, using an MRC can even degrade the
     performance of the destination (with respect to the non-cooperative counterpart) in the case where the relay introduces too
    much decoding noise. In order to extract the best of
    cooperation under any condition
    when DF is assumed, we will present a generalized
    version of the Maximum Likelihood Detector (MLD) originally introduced
    by \cite{laneman-wcnc-2000} and recently re-used by
    \cite{djeumou-isspit-2006}\cite{chen-wireless-2006}.

    \item In Section \ref{sec:af} the MRC was combining, at a given
    cooperation round, the cooperation signal with the last
    recombined signal (from the previous round). It turns out that this assumption really
    complicates the derivation of the optimum detector. In order to
    derive the ML detector, we will suppose that
    the MLD always combines the cooperation signal with the signal
     directly received from the source.

    \item As we have already mentioned, the bandwidth of the signals
    transmitted by the AF-based relays
    has to be equal to the downlink signal bandwidth. When DF is
    assumed, the downlink and cooperation signals can have different
    bandwidths since the relay can use a different modulation from
    the one used by the source. In contrast with the AF case, the
    constraint $\Delta B = B_{DL}$ is therefore relaxed for the DF
    case. In the case of the AF protocol with fixed total bandwidth, the problem of determining the optimum
    number of cooperation exchanges was equivalent to the bandwidth
    allocation problem. Here, the frequency allocation problem
    consists in both, determining the fraction of bandwidth to be allocated
    to the DL channels and determining the number of orthogonal
    sub-bands of the cooperation channel. In this paper, we will not treat
    this issue in its generality since we will only consider the case where the
    downlink bandwidth is fixed. As said earlier, comparing such a cooperative
    system with its non cooperative version ($P_{12}=0$, $P_{21}=0$, $B_C =
    0$) is unarguably unfair in terms of spectral and power
    resources. However, making the assumption $B_{DL} = const.$ has
    two strong advantages: it corresponds to real scenarios engineers have to face with
    and it allows us to keep the
    modulation-coding scheme at the source to be fixed.
\end{enumerate}

As it will be seen, these simplifying assumptions will lead to
results and observations that can provide some insight into the way
of cooperating in practical cases, e.g. a DVB system coupled with a
cellular system. Indeed, for DVB systems the DL signal bandwidth is
typically 20 MHz while receivers in cellular systems have a
bandwidth of a couple of MHz (5 MHz in UMTS systems). Taking into
account the fact that the DF protocol does not impose the DL and the
cooperation signals bandwidths to be equal, it seems to be suited to
the situation taken for illustration.

\subsection{Symmetric and asymmetric cooperations: definitions}
\label{df-asym-sym}

\begin{figure}[!h]
\centering
\includegraphics[angle=-90, width=0.6\columnwidth]{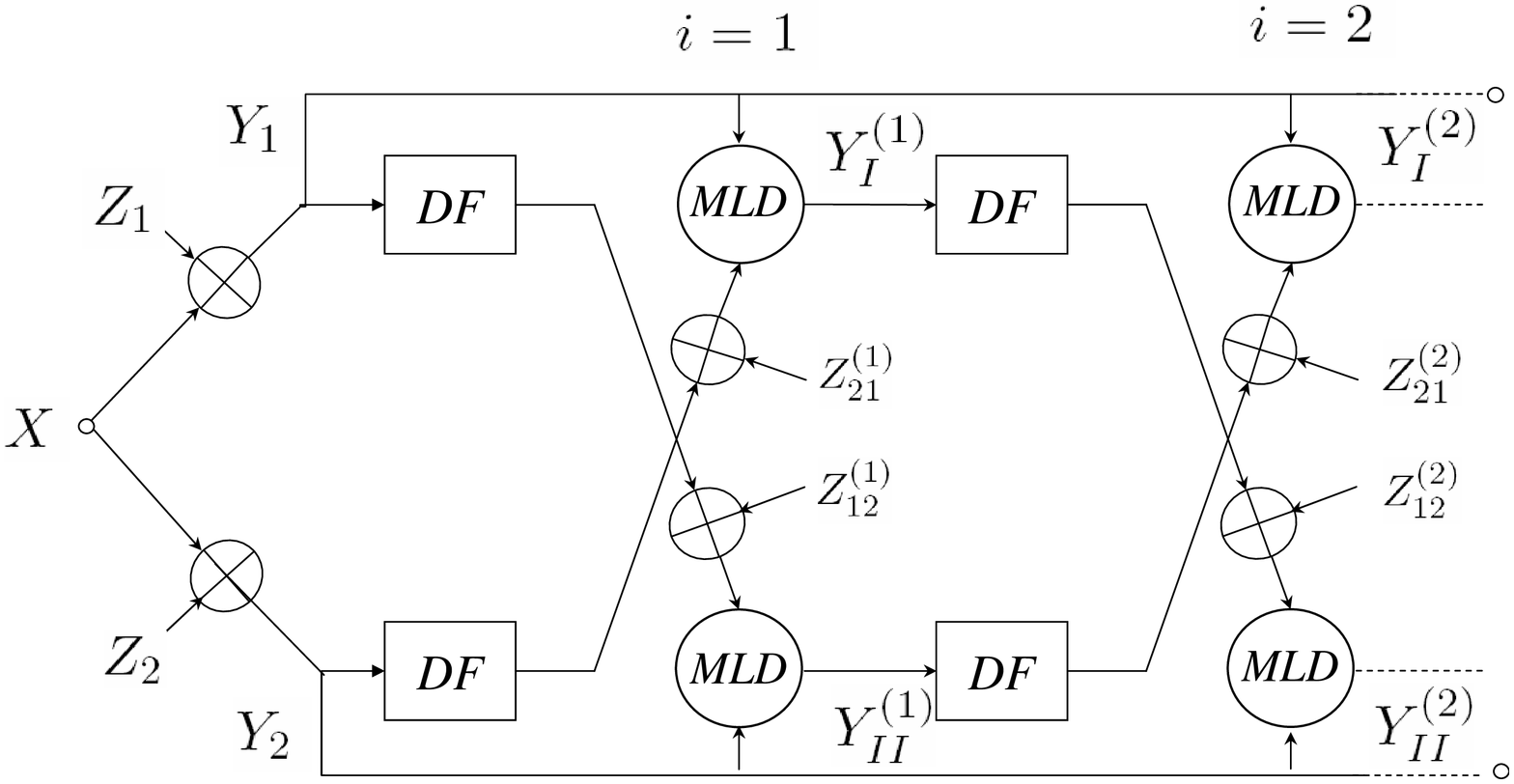}
\caption{\footnotesize{DF-based symmetric cooperation}}
\label{fig-df-symmetric}
\end{figure}

\begin{figure}[!h]
\centering
\includegraphics[angle=-90, width=0.6\columnwidth]{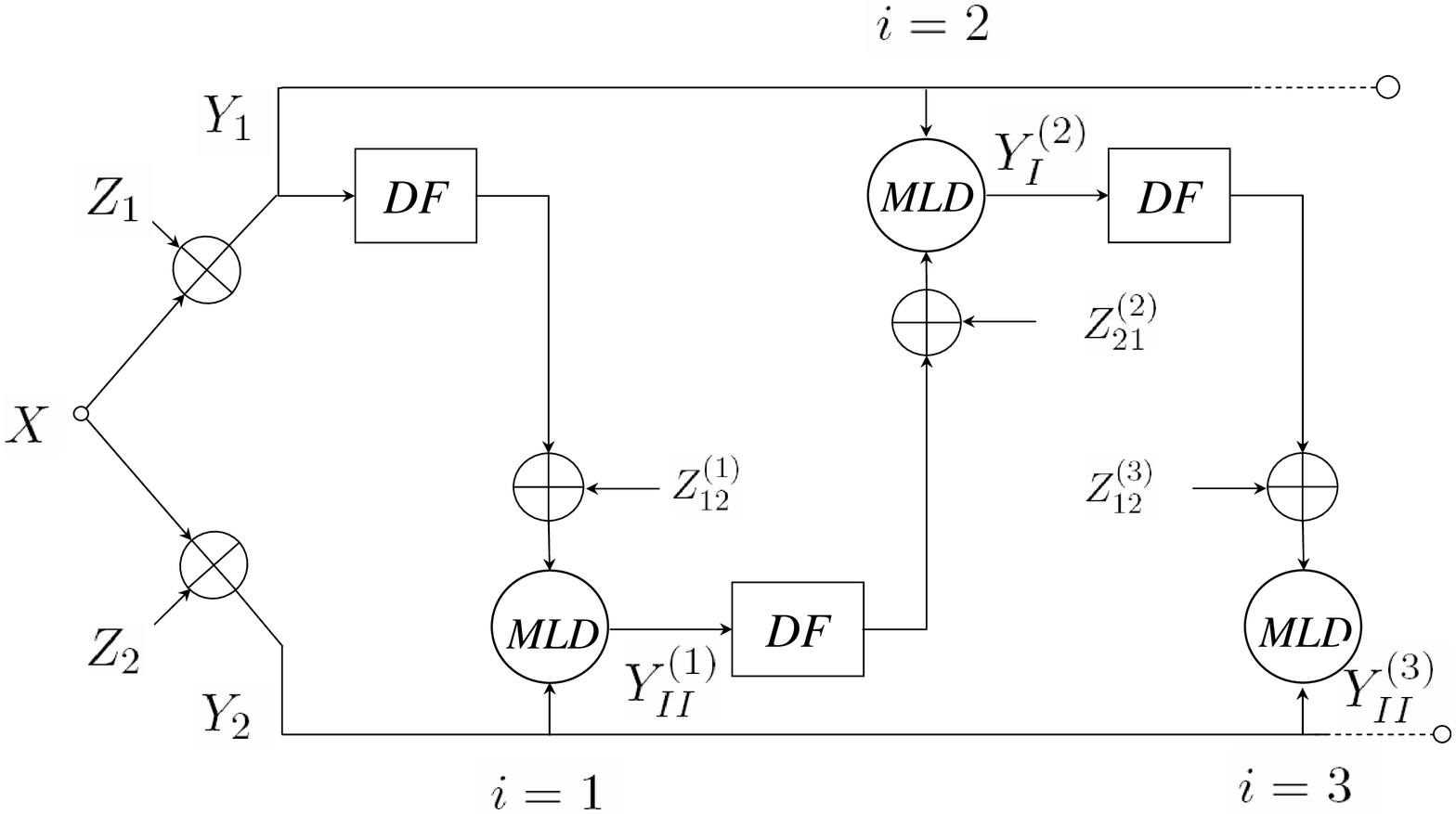}
\caption{\footnotesize{DF-based asymmetric cooperation}}
\label{fig-df-asymmetric}
\end{figure}

Since we have already defined the asymmetric and symmetric cooperations for
the AF protocol, we will just briefly mention the main feature of
the case under investigation. Fig. \ref{fig-df-symmetric} and
\ref{fig-df-asymmetric} define the two corresponding schemes. As
mentioned above, an ML detector is used at the receivers instead of
the MRC. Indeed, the possible presence of decoding noise in the
decoded and forwarded signal makes the equivalent noise at the
receiver non-Gaussian and correlated with the useful signal.
Therefore the equivalent SNR is not always a good performance
criterion. This is why no SNR analysis will be made here. Instead,
we will provide raw BER performance through simulation results.


\subsection{Maximum likelihood detector}
\label{df-combiner}

The purpose of this section is to present a generalized version of
the ML combiner used in
\cite{djeumou-isspit-2006}\cite{laneman-wcnc-2000}\cite{chen-wireless-2006}.
In these works, the authors assumed a binary phase shift keying
modulation at the source and relay, and derived the corresponding
ML detector. Reference \cite{djeumou-isspit-2006} showed that,
under this assumption, the gain provided by the MLD over the MRC
can be significant when the relay has a receive SNR close to (or
less than) the destination SNR, and it is negligible otherwise. In
this paper the reason for extending the MLD of
\cite{djeumou-isspit-2006}\cite{laneman-wcnc-2000}\cite{chen-wireless-2006}
is twofold:
\begin{itemize}
    \item we want the receiver to optimally combine
the signals it receives whatever the noise level at the relay;

    \item it also turns out that the MRC does not seem to be suited for combining signals using
\emph{different constellations} and its derivation is not ready,
perhaps impossible.
\end{itemize}

Before providing the signal model used for the derivation of the
MLD, we consider a special case in order to clearly explain the idea of
compatibility between the modulations used by the source and relay.
Assume that $B_{DL} = B$, $B_C = \frac{B}{4}$, $K=1$ and the source
transmits at the rate of $d = 1 \ \mathrm{bpcu}$. As the relay has
to use the channel twice more often than the source, the relay has
to transmit 4 bpcu in order to send at the same coded bit rate as
the source. For example, if the source and relay implement the same
kind of transmit filters (\emph{e.g.} a root raised-cosine
filter\footnote{For this type of filters the filter bandwidth is
proportional to the symbol rate.}), and the source uses a BPSK
modulation, the relay can use a 16-QAM modulation. In this example
the MLD has to combine one 16-QAM symbol with four BPSK symbols. In
general, the MLD will have to combine $r$ $M_r-$ary symbols from the
relay with $s$ $M_s-$ary symbols from the source, where $r$ and $s$
are linked through the condition of conservation of the coded bit
rate between the input and output of the relay: $r \log_2 M_r = s
\log_2 M_s \triangleq n$.

Without loss of generality assume $K=1$, consider receiver 1 sends a
cooperation signal to receiver 2 and express the signals received by
the latter destination:
\begin{equation}
\label{eq:received-df} \left\{
\begin{array}{ccll}
Y_{2}^{(1)} & = & X^{(1)} + Z_2^{(1)}\\
&\vdots &\\
Y_{2}^{(s)} & = & X^{(s)} + Z_2^{(s)}\\
Y_{12}^{(1)} & = & a_{12}  \epsilon^{(1)}  X_{I}^{(1)} + Z_{12}^{(1)}\\
&\vdots &\\
Y_{12}^{(r)} & = & a_{12}  \epsilon^{(r)}  X_{I}^{(r)} +
Z_{12}^{(r)}
\end{array}
\right.
\end{equation}
where $\forall i \in \{1,...,r\}, \ X^{(i)} \in \{x_1,...,
x_{M_s}\}$, $\forall i \in \{1,...,r\}, \ X_{I}^{(i)} \in
\{x_{I,1},..., x_{I,M_r}\}$ and the random variables
$\epsilon^{(i)}$ model the decoding noise introduced by the relay.
For example, when the relay uses a QPSK modulation,
$\epsilon^{(i)} \in \{1, e^{j\frac{\pi}{2}}, e^{j\pi}, e^{j\frac{3
\pi}{2}}\}$. Now, in order to express the likelihood at receiver
2, we introduce the following notations: $\ul{Y}_{2} =
(Y_{2}^{(1)},...,Y_{2}^{(s)} )$, $\ul{Y}_{12} =
(Y_{12}^{(1)},...,Y_{12}^{(r)} )$, $\ul{b} = ( b_1,...,b_n )$
denotes the vector of coded bits associated with the
\emph{ordered} vector of symbols $\ul{X} = (X^{(1)}, ..., X^{(r)}
)$. We want to express the likelihood $p_{ML} =
p(\ul{y}_{2},\ul{y}_{12} \ | \ \ul{b})$. We have
\begin{equation}
p_{ML}   = p\left(\ul{y_{2}},\ul{y_{12}} \ | \ \ul{b}\right) \stackrel{(a)}{=}
 p\left(\ul{y}_{2},\ul{y}_{12} \ | \ \ul{x}\right) \nonumber\\
                \stackrel{(b)}{=} p\left(\ul{y}_{2} \ | \ \ul{x}\right) \ p\left(\ul{y}_{12} \ | \ \ul{x}\right)
                \label{eq:mv}
\end{equation}
where \\
(a) there is a one-to-one mapping between $\ul{X}$ and $\ul{b}$;\\
(b) the noises of the downlink and the cooperative channels are
independent.

Denoting $N_2 = n_2 B_{DL}$, the first term of the product in
equation (\ref{eq:mv}) expresses as
\begin{eqnarray}
p \left(\ul{y_{2}} \ | \ \ul{x}\right)  &=& p\left(y_2^{(1)},...,y_2^{(r)} \ | \ x^{(1)},...,x^{(r)} \right) \nonumber\\
                &=& \prod_{i=1}^{r} p\left(y_2^{(i)} \ | \ x^{(i)} \right) \nonumber\\
                &=& \prod_{i=1}^{r} \frac{1}{\pi N_2}\exp\left(-\frac{\left|y_2^{(i)} - x^{(i)}
                \right|^2}{N_2}\right).
                \label{eq:mv1}
\end{eqnarray}
By denoting $N_{12} = n_{12} \Delta B$ the second term can be
expanded as follows
\begin{eqnarray}
p \left(\ul{y_{12}} \ | \ \ul{x}\right) &=& p \left(\ul{y_{12}} \ | \ \ul{b}\right) \nonumber\\
                &=& p\left(y_{12}^{(1)},...,y_{12}^{(s)} \ | \ b_1,...,b_n \right) \nonumber\\
                &\stackrel{(c)}{=}& \prod_{i=1}^{s} \ p\left(y_{12}^{(i)}
                \ | \ b_{(i-1)\log_2 M_s+1},...,b_{i \log_2 M_s} \right) \nonumber\\
                &=& \prod_{i=1}^{s} \ p\left(y_{12}^{(i)} \ | \ x_{I}^{(i)} \right) \nonumber\\
                &\stackrel{(d)}{=}& \prod_{i=1}^{s} \ \sum_{j=1}^{M_s} \
                 \mathrm{Pr}\left[\epsilon^{(i)} = \epsilon^{(i)}_j \left| x_{I}^{(i)} \right. \right] p\left(y_{12}^{(i)} \ | \ x_{I}^{(i)}, \epsilon^{(i)} = \epsilon^{(i)}_j \right) \nonumber\\
                &=& \prod_{i=1}^{s} \ \sum_{j=1}^{M_s} \
                \mathrm{Pr}\left[\epsilon^{(i)} = \epsilon^{(i)}_j  \left| x_{I}^{(i)} \right.  \right] \
                \frac{1}{\pi N_{12}}\exp\left(-\frac{\left|y_{12}^{(i)}
                - \epsilon^{(i)}_j x_{I}^{(i)} \right|^2}{N_{12}}\right)
                \label{eq:mv2}
\end{eqnarray}
with\\
(c) given $x_I^{(i)}$, the signal $y_{12}^{(i)}$ is independent of
$x_I^{(j)}$ for $j \neq i$; remind that $x_I^{(i)}$
is associated with $(b_{(i-1)\log_2 M_s+1},...,b_{i \log_2 M_s})$.\\
(d) is obtained by marginalizing over $\epsilon^{(i)}$.

As in \cite{murthy-ncc-2005}, we want to express the log likelihood
ratio associated with a given coded bit as a function of the
likelihood expressed above. To this end, let us define the sets:
$\mc{B}_{i}^{(n)}(k) = \{\ul{b} \in \{0,1\}^n, \ b_k = i$ with $i =
0$ or $i=1$. The coded bits $b_k$ being equiprobable we have:
\begin{eqnarray}
LLR\left(b_{k}\right)   &\triangleq&
\frac{p\left(\ul{y}_{2},\ul{y}_{12} \ | \ b_{k} = 1\right)
}{p\left(\ul{y}_{2},\ul{y}_{12} \ | \ b_{k} = 0\right)} \nonumber \\
                                            &=& \frac{\ds{\sum_{\ul{b} \in \mc{B}^{(n)}_{1}(k) } }
                                             p\left(\ul{y}_{2},\ul{y}_{12} \ | \ \ul{b}
                                              \right)}{\ds{\sum_{\ul{b} \in \mc{B}^{(n)}_{0}(k) }}
                                              p\left(\ul{y}_{2},\ul{y}_{12} \ | \ \ul{b} \right)} \nonumber\\
                                            &=& \frac{\ds{\sum_{\ul{x} \in \mc{X}^{k}_{1}} }
                                             \ p\left(\ul{y}_{2},\ul{y}_{12} \ | \ \ul{x}
                                              \right)}{\ds{\sum_{\ul{x} \in \mc{X}^{k}_{0}}}
                                              \ p\left(\ul{y}_{2},\ul{y}_{12} \ | \ \ul{x}
                                              \right)}.
                \label{eq:llr1}
\end{eqnarray}
This LLR can be either used to make a decision on the bits sent by
the source or re-used as a soft information by a stage following the
combiner. As we restrict our attention to the raw BER for our
performance study, we will not consider the way of using this LLR by
the channel decoder for example.

\section{Experimental analysis}
\label{sec:experimental}

\subsection{System performance criteria}
\label{sec:perf-crit}

In order to compare the different cooperation schemes, suited
\emph{system} performance criteria have to be selected. By way of an
example, if we fix the information rate/spectral efficiency at the
transmitter and obtain the pair of BERs $(BER_1, BER_2)$ for the
coding scheme $\mc{C}$ and $(BER_1', BER_2')$ for the coding scheme
$\mc{C}'$, with $BER_1 < BER_1'$ and $BER_2
> BER_2'$, one cannot easily conclude, which shows the importance of
using a system performance metric. From now on, we will denote by
$K$ the number of cooperation exchanges with $K$ equals $K_a$ or
$2K_s$ depending on the cooperation scheme. In order to compare the
different cooperation strategies, we propose four performance
criteria (eq. (\ref{eq:crit1})-(\ref{eq:crit4})). All the
performance criteria can be used to evaluate the performance of
the system for both relaying protocols but the performance criterion
1) is less meaningful for the DF protocol since the channel input is
not Gaussian in our context.
\begin{enumerate}
\item
 \begin{equation}
    \label{eq:crit1}
    R_{AF}^{(K)}  = B_{DL}
    \min \left\{ \log \left( 1 + \rho_{I}^{(K)}  \right),
     \log  \left( 1 + \rho_{II}^{(K)}  \right) \right\}
 \end{equation}
      where $\rho_{I}^{(K)}$,
      $\rho_{II}^{(K)}$ are the SNRs at the end of the
      cooperation procedure. One can notice that $R_{AF}^{(K)}$ represents
      the maximum information rate possible for a reliable
      transmission achieved by the AF-based
      cooperation schemes and a Gaussian channel input.

 \item
\begin{equation}
\label{eq:crit2}
    P_{e,max}^{(K)} = \max \left\{ P_{e,I}^{(K)}, P_{e,II}^{(K)}
\right\}
\end{equation}
 where $P_{e,I}^{(K)}$ and $P_{e,II}^{(K)}$
are the raw BERs at the combiner (\emph{i.e.} the MRC for the AF
protocol, the MLD for the DF protocol)
    outputs at the end of the cooperation procedure. This
    criterion is useful in a broadcasting system for which one wants every user
    to have a minimum transmission quality, which requires $P_{e,max}^{(K)} \leq
    P_{e,0} $ where $P_{e,0}$ is the minimum quality target.
\item
    \begin{equation}
    \label{eq:crit3}
     P_{e,sum}^{(K)} =  P_{e,I}^{(K)} +
    P_{e,II}^{(K)}.
    \end{equation}
    This
    criterion gives an image of the average transmission quality
    of the broadcasting system and serves as an upper bound for the
    performance criterion given just below. Although this criterion
    does not translate the variance of the qualities of the
    different communications it has the merit to be simple, which is
    the reason why many works assumed this criterion (see \emph{e.g.} \cite{frenger-frames-1999}\cite{benvenuto-pimrc-2004}).
 \item
    \begin{equation}
    \label{eq:crit4}
    P_{e,max}^{(K)} \leq P_{e,sys}^{(K)} \leq P_{e,sum}^{(K)}.
    \end{equation}
    The quantity $P_{e,sys}^{(K)}$ is the system probability of errors $P_{e,sys}^{(K)}$, which is
    defined as the probability that receiver 1 or (inclusive
    or) receiver 2 makes a decision error. This probability is generally
    not easy to explicit but can be bounded by using the criteria (\ref{eq:crit2})
    and (\ref{eq:crit3}). As a comment, note that the Shannon capacity of the channel under
consideration is
    precisely defined with respect to the system error
    probability, which means that communicating at a rate less
    than the capacity insures the existence of a code such that
    $P_{e,sys}^{K^*} \rightarrow 0$. It is therefore the
    criterion to be considered to assess the sub-optimality of a given channel coding
    scheme in the CBC w.r.t its Shannon limit.
\end{enumerate}

\subsection{Simulation results for the AF protocol}
\label{sec:simuls_af}

On all figures $R^K_{AF}$ and $R^K_{AF}-DL$ denote the achievable
rates with the strategy $\mc{S}_1$ and the strategy $\mc{S}_2$
respectively.

\begin{figure}[h!]
  \begin{center}
        \includegraphics[height=0.40\textwidth, width=0.49\textwidth]
    {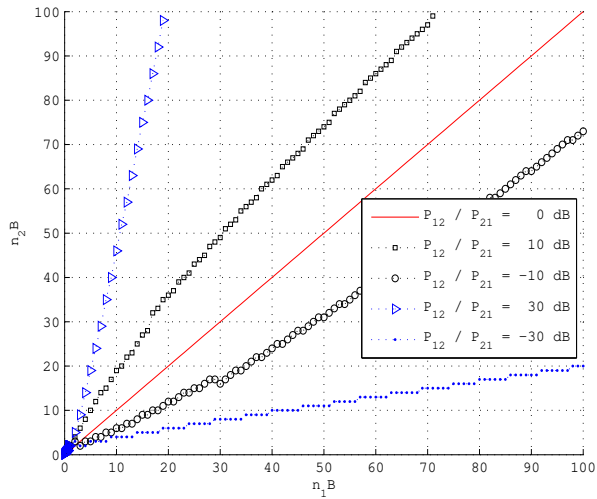}
  \end{center}
  \caption{Decision regions: who starts cooperating
first?}
  \label{who-starts}
\end{figure}

\emph{Asymmetric cooperation: Which user should start cooperating
first?}\\
For both strategies $\mc{S}_1$  and $\mc{S}_2$, Figure
\ref{who-starts} represents the plane $(n_1, n_2)$ with linear
scales: $n_1 \in [10^{-2},10^{2}]$, $n_2 \in [10^{-2},10^{2}]$. For
different ratios\\ $\frac{P_{12}}{P_{21}} \in \{-30 \ \mathrm{dB},
-10 \ \mathrm{dB}, 0 \ \mathrm{dB},10 \ \mathrm{dB},30 \ \mathrm{dB}
\}$. The different curves delimit the decision regions that allow us
to determine the best cooperation order in terms of information rate
for the five values of the ratio $\frac{P_{12}}{P_{21}}$. When the
pair $(n_1,n_2)$ is above the line, receiver 1 has to start first
and conversely. We see that both the DL and cooperation SNRs have to
be considered to optimize the overall performance. In a cellular
system for instance, the cooperation powers can be quite close (a
given fraction of the mobile transmit power), which would make the
cooperation order less critical.

\emph{Comparison between the strategies $\mc{S}_1$  and $\mc{S}_2$.}

\begin{figure}[h!]
  \begin{center}
    \subfigure[]{\label{fig:comp-AF-vs-new-high-coop}\includegraphics[height=0.40\textwidth, width=0.49\textwidth]
    {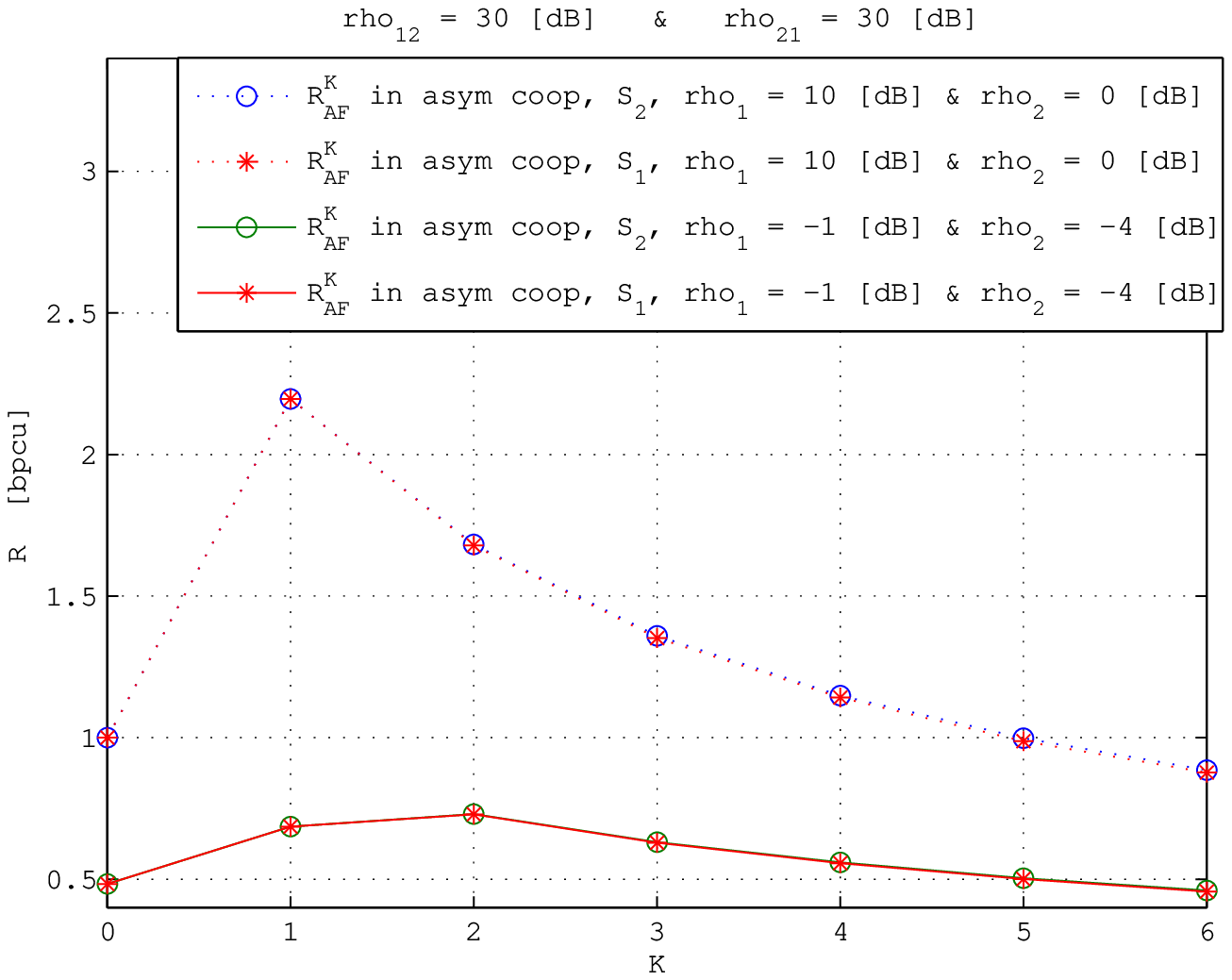}}
    \subfigure[]{\label{fig:comp-AF-vs-new-med-coop}\includegraphics[height=0.40\textwidth, width=0.49\textwidth]
    {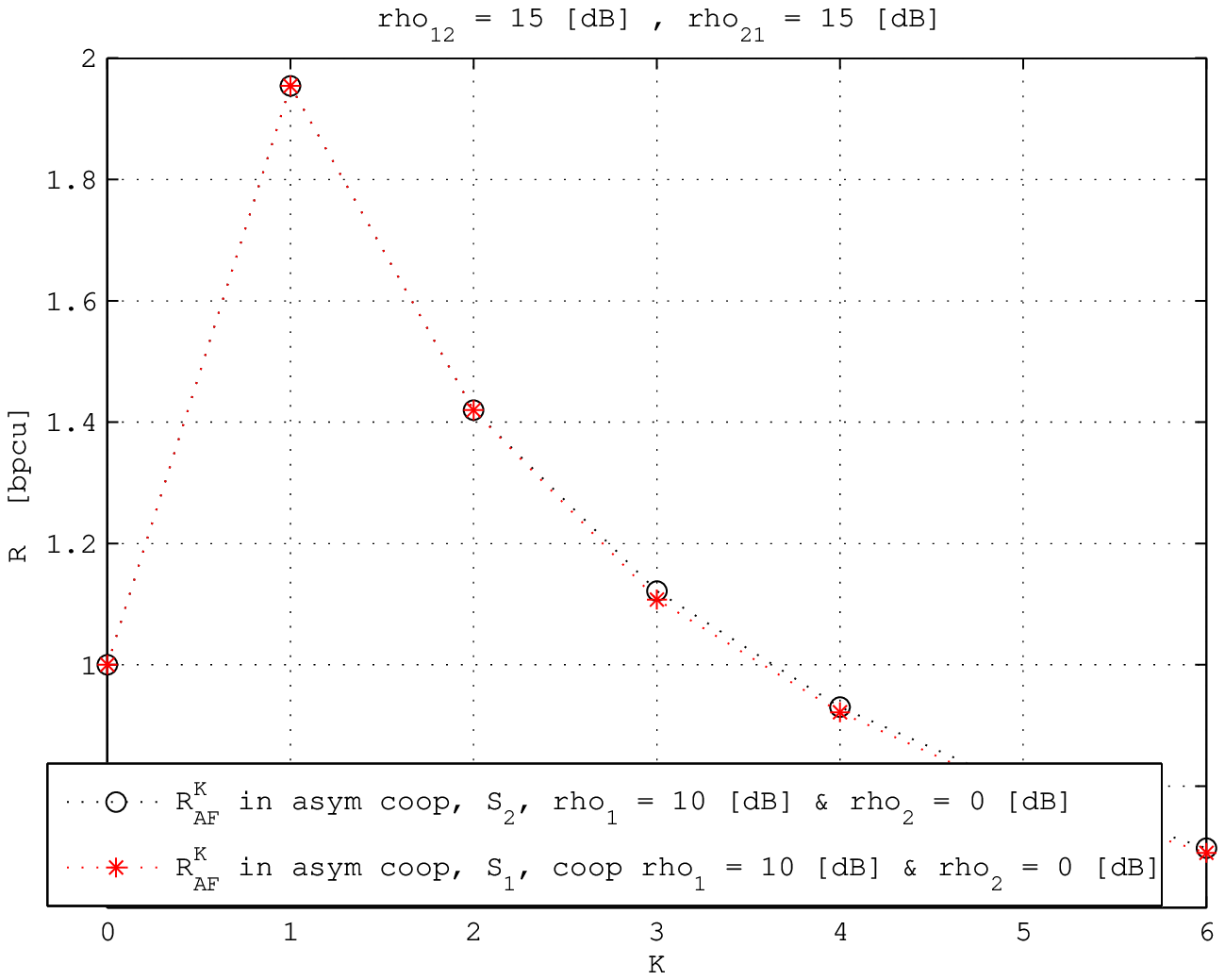}}
  \end{center}
  \caption{Achievable rate vs number of cooperation
exchanges for both strategies $\mc{S}_1$ and $\mc{S}_2$ when the
total bandwidth is fixed at high cooperative regime\ (a)  \ or
medium cooperative regime \ (b).}
\end{figure}

We first consider the case of a constant the global bandwidth. We
look at three different SNR scenarios:\\
$(\frac{P}{n_1 B}, \frac{P}{n_2 B}, \frac{P_{12}}{n_{12} B},
\frac{P_{21}}{n_{21} B}, ) = (10 \ \mathrm{dB}, 0 \ \mathrm{dB}, 30
\ \mathrm{dB}, 30 \ \mathrm{dB})$,  $(\frac{P}{n_1 B}, \frac{P}{n_2
B}, \frac{P_{12}}{n_{12} B}, \frac{P_{21}}{n_{21} B}, ) = (-1 \
\mathrm{dB}, -4 \ \mathrm{dB}, 30 \ \mathrm{dB}, 30 \ \mathrm{dB})$
and $(\frac{P}{n_1 B}, \frac{P}{n_2 B}, \frac{P_{12}}{n_{12} B},
\frac{P_{21}}{n_{21} B}, ) = (10 \ \mathrm{dB}, 0 \ \mathrm{dB}, 15\
\mathrm{dB}, 15 \ \mathrm{dB})$. Figure
\ref{fig:comp-AF-vs-new-high-coop} and
\ref{fig:comp-AF-vs-new-med-coop} represent the performances of both
strategies $\mc{S}_1$  and $\mc{S}_2$ where the asymmetric cooperation
is considered.
Both strategies have approximately the same performance but the
strategy $\mc{S}_2$ can perform better than the strategy $\mc{S}_1$
for great values of $K$ ($K > 2$, see Figure
\ref{fig:comp-AF-vs-new-med-coop}). Since the optimum is obtained in general at
low values of $K$ ($K \leq 2$), we can conclude that both
strategies $\mc{S}_1$ and $\mc{S}_2$ have similar performances in
asymmetric cooperation. We have also observed that this conclusion remains
valid when the symmetric cooperation is considered.

\begin{figure}[h!]
  \begin{center}
    \includegraphics[height=0.40\textwidth, width=0.49\textwidth]
    {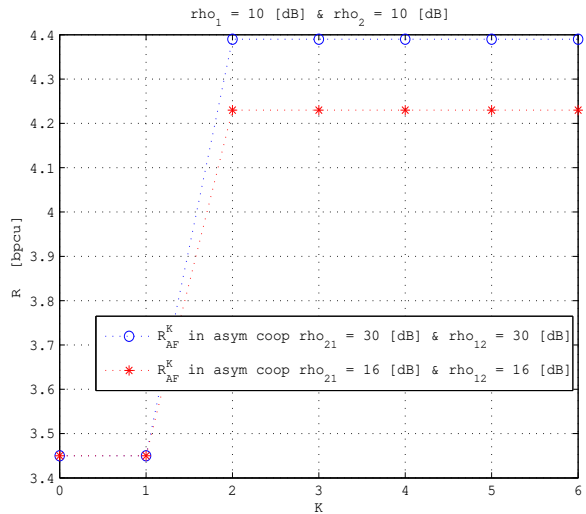}
  \end{center}
  \caption{Achievable rate vs number of cooperation
exchanges for the strategy $\mc{S}_2$ when the downlink bandwidth is
fixed for medium and high cooperative regimes.}
  \label{fig:new-med-and-high-coop}
\end{figure}

Now, we consider the case where the DL bandwidth is constant. In
Figure \ref{fig:new-med-and-high-coop} can consider two different scenarios for the $\mc{S}_2$ for the asymmetric cooperation case:\\
$(\frac{P}{n_1 B}, \frac{P}{n_2 B}, \frac{P_{12}}{n_{12} B},
\frac{P_{21}}{n_{21} B} ) = (10 \ \mathrm{dB}, 10 \ \mathrm{dB}, 30
\ \mathrm{dB}, 30 \ \mathrm{dB})$ and $(\frac{P}{n_1 B},
\frac{P}{n_2 B}, \frac{P_{12}}{n_{12} B}, \frac{P_{21}}{n_{21} B}, )
= (10 \ \mathrm{dB}, 10 \ \mathrm{dB}, 16 \ \mathrm{dB}, 16 \
\mathrm{dB})$. We observe that the SIMO bound is rapidly attained ($K=2$). Also we
have observed that, for the strategy $\mc{S}_2$,
the symmetric and asymmetric cooperations perform the same. This means that for the symmetric case also the SIMO bound is attained for $K_s=1$. In the following paragraph we will compare these results with the results obtained with the stategy $\mc{S}_1$ in the same setup ( DL bandwidth constant).

\begin{figure}[h!]
  \begin{center}
    \subfigure[]{\label{sym-vs-asym}\includegraphics[height=0.40\textwidth, width=0.49\textwidth]
    {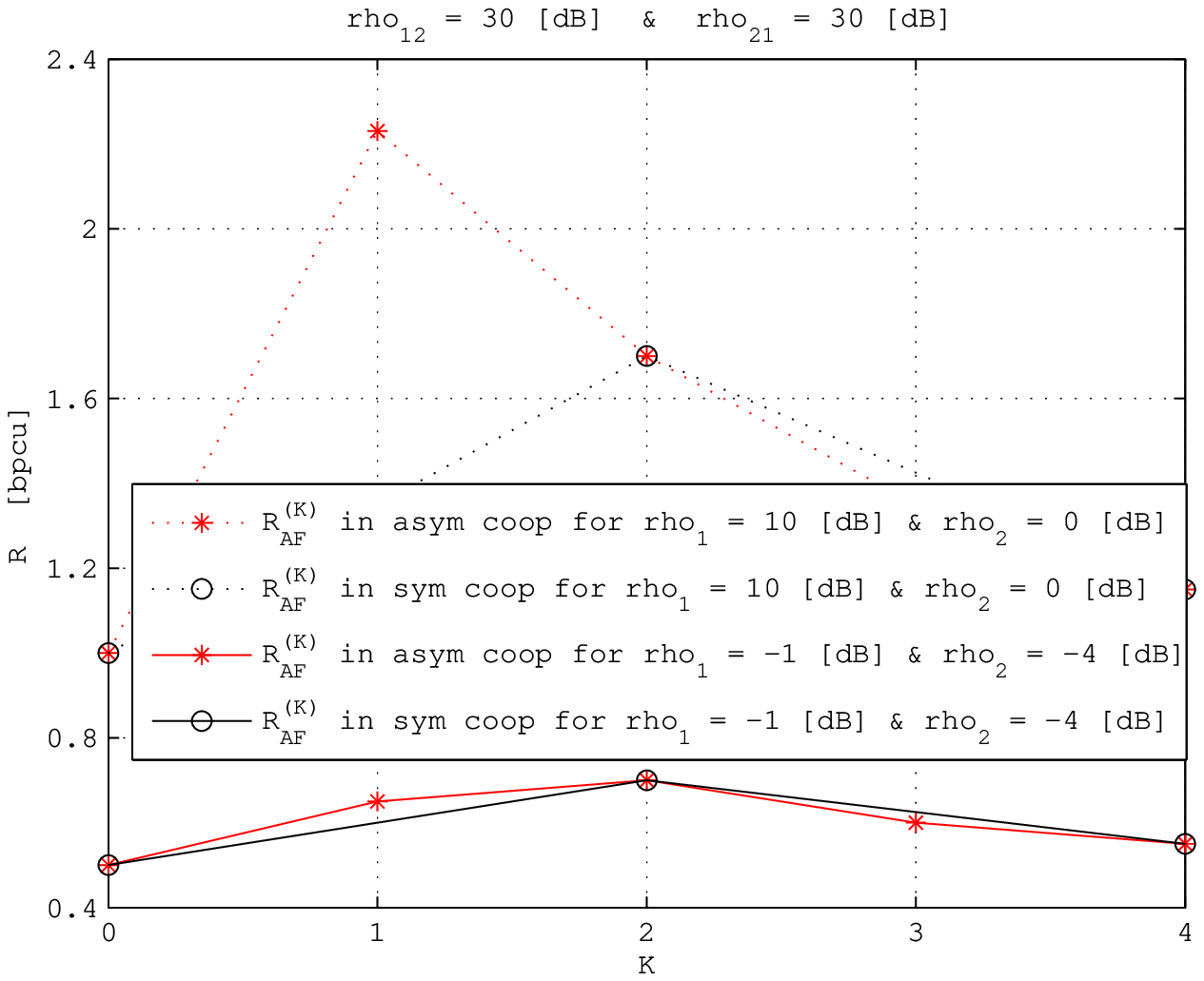}}
    \subfigure[]{\label{sym-vs-asym-2}\includegraphics[height=0.40\textwidth, width=0.49\textwidth]
    {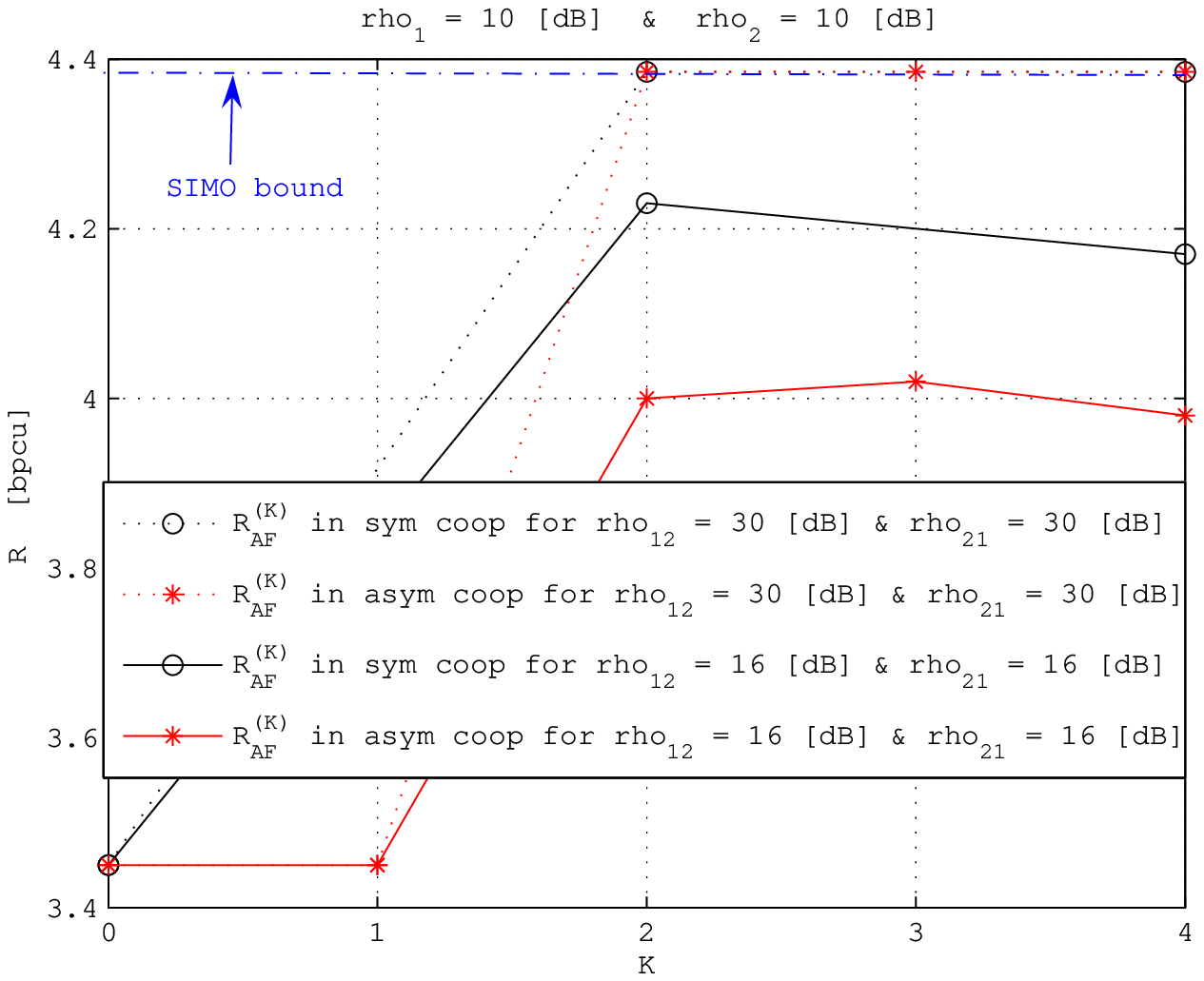}}
  \end{center}
  \caption{Achievable rate vs number of cooperation
exchanges for the strategy $\mc{S}_1$ when the total bandwidth is fixed \ (a)  \ or when the
downlink bandwidth is fixed \ (b).}
\end{figure}

\emph{Asymmetric cooperation vs symmetric cooperation for the
strategy $\mc{S}_1$.} First we assume the \emph{total} bandwidth to
be limited. Figure \ref{sym-vs-asym} represents the information rate
as a function of the number of cooperation exchanges for the
asymmetric  and symmetric
 cases for two different scenarios:\\
$(\frac{P}{n_1 B}, \frac{P}{n_2 B}, \frac{P_{12}}{n_{12} B},
\frac{P_{21}}{n_{21} B}, ) = (10 \ \mathrm{dB}, 0 \ \mathrm{dB}, 30
\ \mathrm{dB}, 30 \ \mathrm{dB})$ and  $(\frac{P}{n_1 B},
\frac{P}{n_2 B}, \frac{P_{12}}{n_{12} B}, \frac{P_{21}}{n_{21} B}, )
= (-1 \ \mathrm{dB}, -4 \ \mathrm{dB}, 30 \ \mathrm{dB}, 30 \
\mathrm{dB})$. It can be seen that the rate always decreases for $K
\geq 2$. This is not surprising since a system with $K > 2$ is a
special case of the system for which $K=2$. However note that the
system with $K=2$ is not a special of the system $K = 0$ or $K =1$,
which means that cooperating can compensate for the performance loss
due do orthogonalizing the DL channel. We also see that the
asymmetric system performs better than its symmetric counterpart. We
observed from other simulations not presented here that most of the
cooperation benefits are captured with one cooperation exchange. In
contrast with the discrete CBC with a conference channel
\cite{draper-allerton-2003}\cite{khalili-allerton-2006}\cite{dabora-it-2006}
we see that the performance can decrease with $K$. Now we look at
two
scenarios where the \emph{downlink} bandwidth is fixed (Figure \ref{sym-vs-asym-2}):\\
$(\frac{P}{n_1 B}, \frac{P}{n_2 B}, \frac{P_{12}}{n_{12} B},
\frac{P_{21}}{n_{21} B} ) = (10 \ \mathrm{dB}, 10 \ \mathrm{dB}, 30
\ \mathrm{dB}, 30 \ \mathrm{dB})$ and $(\frac{P}{n_1 B},
\frac{P}{n_2 B}, \frac{P_{12}}{n_{12} B}, \frac{P_{21}}{n_{21} B}, )
= (10 \ \mathrm{dB}, 10 \ \mathrm{dB}, 16 \ \mathrm{dB}, 16 \
\mathrm{dB})$. We see that in the high cooperation regime the SIMO
bound is rapidly attained; that is for $K=2$. When less cooperation
powers are available the performance still decreases with $K$. This
time this is not due to the orthogonalization loss but to the fact
that the cooperation power per exchange decreases in $ \sim
\frac{1}{K}$ whereas the gain brought by increasing the number of
recombinations increases slowly. Note that now the symmetric system
performs better than the asymmetric one because nothing is lost in
terms of bandwidth by increasing $K$ (while for the case where the
total bandwidth was limited the DL bandwidth was decreasing
according to propositions \ref{prop-symm} and \ref{prop-asymm}).

Comparing the results of the strategies $\mc{S}_1$ and $\mc{S}_2$ when the DL bandwidth is constant ( Figures
\ref{sym-vs-asym-2} and \ref{fig:new-med-and-high-coop}), we have observed that both strategies perform identically for the
symmetric cooperation when there is enough power available for the cooperation (high cooperative regime). If the cooperative
power is reduced, the strategy $\mc{S}_2$ will perform better than the strategy $\mc{S}_1$, starting from $K_s = 2$ (equivalent
to $K=2$). In fact, using the strategy $\mc{S}_1$, during the second exchange round, the receiver acting as relay is waisting a
part of the limited available power to send to the other receiver a signal that it has already received on the downlink
channel. Thus, the expected power gain from the cooperation is limited w.r.t. the strategy $\mc{S}_2$, where the receiver
acting as relay uses all of the available power to send the signal needed to increase the equivalent SNR. However, since the
optimal performance is obtained for $K_s = 1$, we can conclude that both strategies have the same performance for symmetric
cooperation case.

For the strategy $\mc{S}_2$ the symmetric and asymmetric cooperation schemes perform identically. This is also the case for the
strategy $\mc{S}_1$ but only when the high cooperative regime is assumed. For the strategy $\mc{S}_2$, the achievable rates
remain constant after $K_s = 1$, whatever the cooperation power level. For the strategy $\mc{S}_1$, if the cooperation powers
are limited, the symmetric cooperation case outperforms its asymmetric counterpart. Also, for the asymmetric cooperation with
limited cooperation powers, the strategy $\mc{S}_2$ performs better than the strategy $\mc{S}_1$ even at the optimum number of
cooperation rounds $K^*$.

\emph{Influence of the performance criterion and BER analysis.}

\begin{figure}[h!]
  \begin{center}
    \subfigure[]{\label{ber}\includegraphics[height=0.40\textwidth, width=0.49\textwidth]{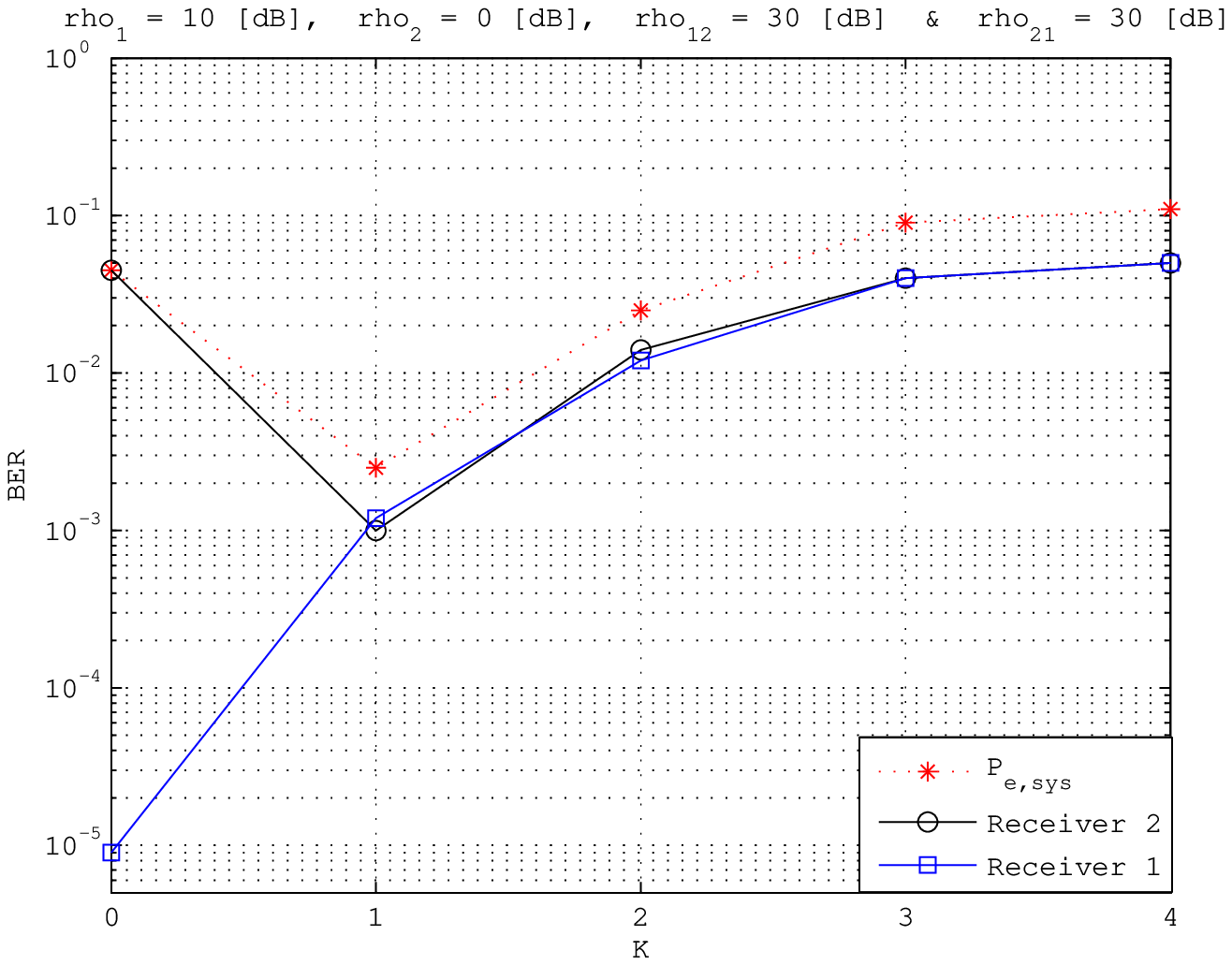}}
        \subfigure[]{\label{ber-all-medium}\includegraphics[height=0.40\textwidth, width=0.49\textwidth]{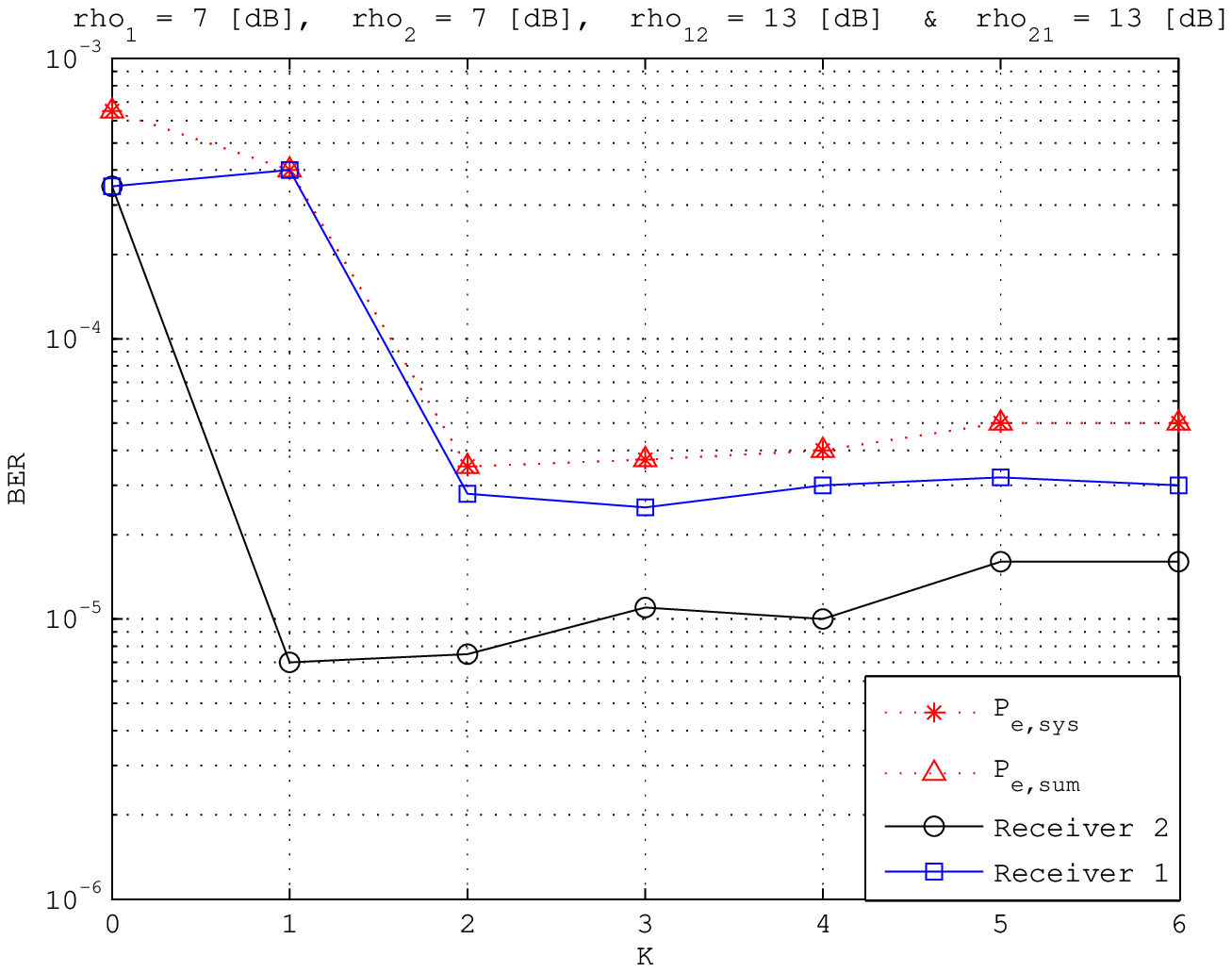}}
  \end{center}
  \caption{System BER vs number of cooperation exchanges for strategy $\mc{S}_1$ and the asymmetric cooperation case when the total bandwidth is fixed \ (a)  \ or when the downlink bandwidth is fixed \ (b).}
\end{figure}

In the simulations results presented so far we have implicitly assumed the channel input and
relay outputs to be Gaussian, which allowed us to provide an
achievable rate for the channel under investigation. In the following part of the section we will assume finite modulations (QAM modulations). It
turns out that the observations made for the information theoretic
transmission rate are generally confirmed by the raw BER analysis
and under the QAM assumption. This fact is illustrated in Figures \ref{ber} and
\ref{ber-all-medium}. In both figures the asymmetric cooperation case and the strategy $\mc{S}_1$ is assumed. Also, the first figure corresponds to
Assumption $\mc{H}_1$ while the second one is based on Assumption
$\mc{H}_1$. The system BER is minimized for $K=1$ or $K=2$ whatever
the assumption on spectral resources.

\subsection{Simulation results for the DF protocol}
\label{sec:simuls_df}

In the case of the DF protocol, we always assume that only the
downlink bandwidth is fixed. As a consequence the total bandwidth increases with $K$ (see Assumption $\mc{H}_2$ in sec. \ref{sec:system-model}).\\

\emph{Asymmetric cooperation vs symmetric cooperation.}

\begin{figure}[h!]
  \begin{center}
    \subfigure[]{\label{sym-asym-max}\includegraphics[height=0.40\textwidth, width=0.49\textwidth]
    {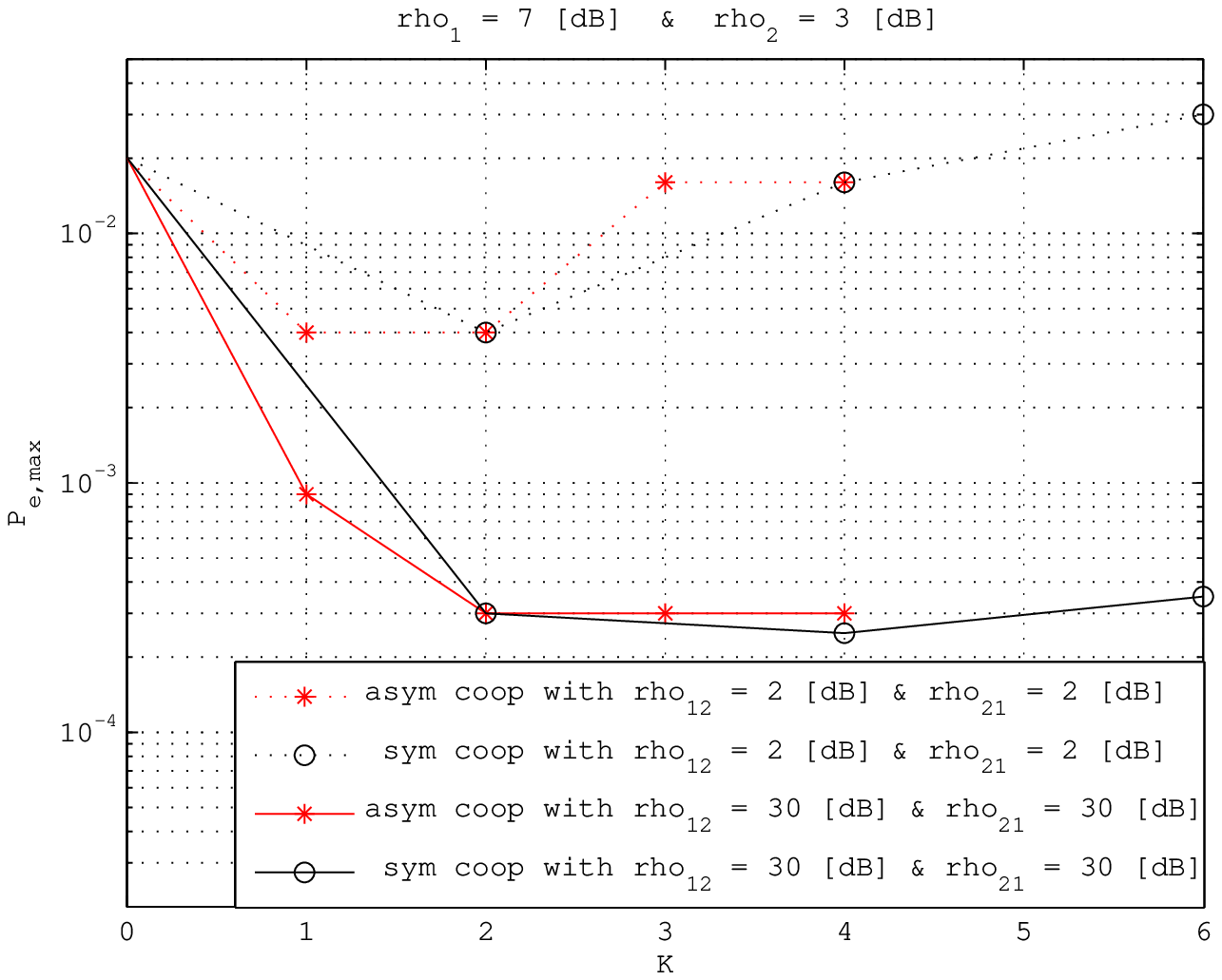}}
        \subfigure[]{\label{sym-asym-sys}\includegraphics[height=0.40\textwidth, width=0.49\textwidth]
    {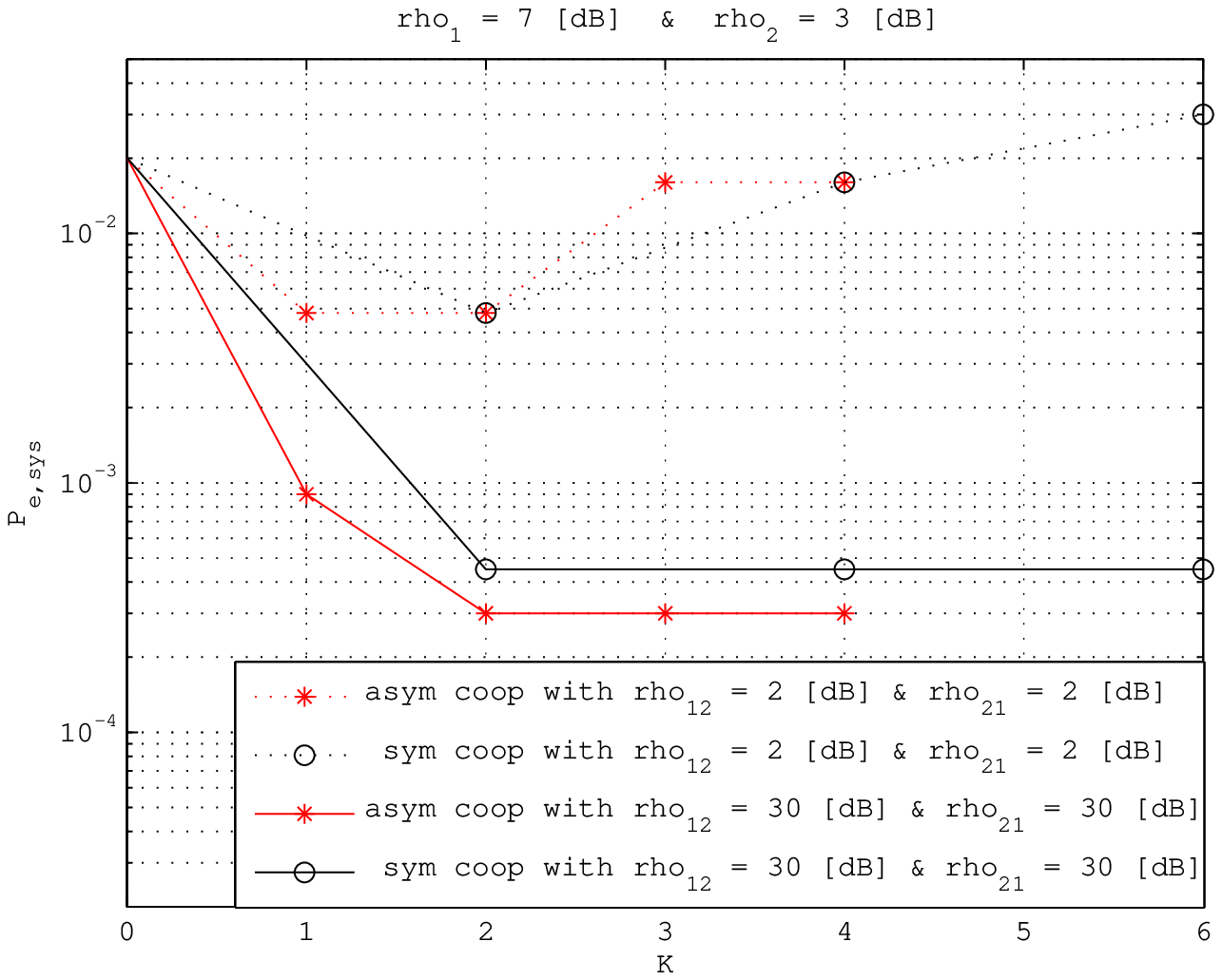}}
  \end{center}
  \caption{Performance vs number of cooperation exchanges in term of $P_{e,sys}$  \ (a) and
  $P_{e,max}$ \ (b).}
\end{figure}

We always assume QAM modulations at the source and relays and we do
take into account the possible presence of channel coders at the
source and relay. We consider two different scenarios: a high
cooperative regime with $(\frac{P}{n_1 B}, \frac{P}{n_2 B},
\frac{P_{12}}{n_{12} B}, \frac{P_{21}}{n_{21} B}, ) = (7 \
\mathrm{dB}, 3 \ \mathrm{dB}, 30 \ \mathrm{dB}, 30 \ \mathrm{dB})$
and a low cooperative regime with $(\frac{P}{n_1 B}, \frac{P}{n_2
B}, \frac{P_{12}}{n_{12} B}, \frac{P_{21}}{n_{21} B}, ) = (7 \
\mathrm{dB}, 3 \ \mathrm{dB}, 2 \ \mathrm{dB}, 2 \ \mathrm{dB})$. We
use a 4-QAM modulation for any transmission at the source and at the
relay. We only consider the uncoded case but the performance
analysis can be extended to coded case, at least for hard input
decoders. We assume that receiver 1 starts sending a cooperation in
the asymmetric cooperation case. Figures \ref{sym-asym-sys} and
\ref{sym-asym-max} show the system performance as a function of the
number of cooperation exchanges for performance criteria 2. and 4.
respectively. In the low cooperation regime, symmetric and
asymmetric cooperations perform similarly. In the high cooperation
regime, the asymmetric cooperation performs slightly better for
$P_{e,sys}$ and conversely for $P_{e,max}$. Other simulations, which
will not provided here for keeping the number of figures reasonable,
show that the performance of asymmetric cooperation is generally
better than that of its symmetric counterpart, whatever the
performance criterion under consideration. In contrast with the AF
case it is more difficult to determine analytically which receiver
has to start cooperation in the first place. This means that, in
practice, this information has to be sent to the receivers.
Otherwise, the symmetric cooperation has to be used.

\emph{Asymmetric cooperation: influence of both number of
cooperation exchanges and combining scheme.}

\begin{figure}[h!]
  \begin{center}
    \subfigure[]{\label{asym_low}\includegraphics[height=0.40\textwidth, width=0.49\textwidth]
    {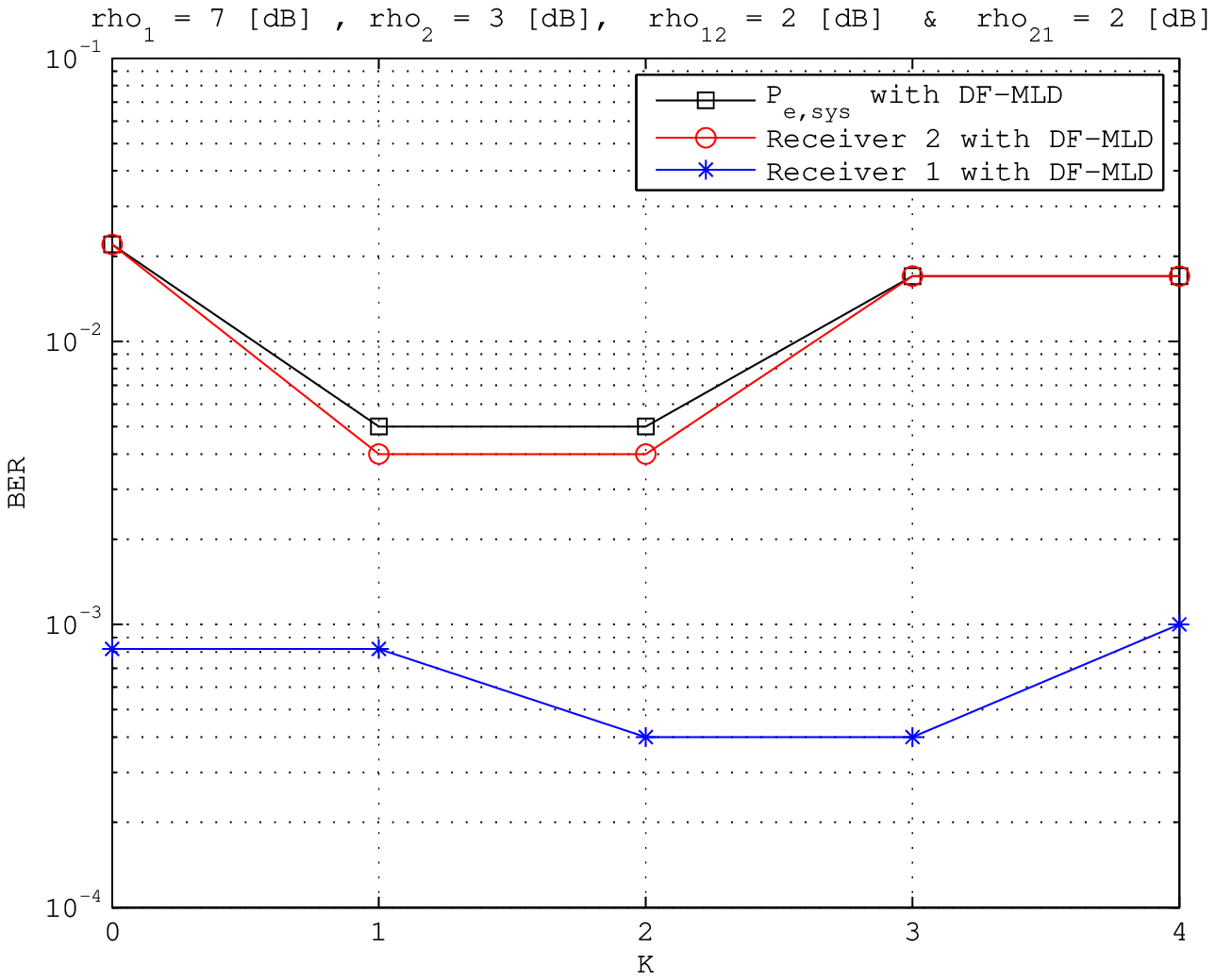}}
        \subfigure[]{\label{asym_high}\includegraphics[height=0.40\textwidth, width=0.49\textwidth]
    {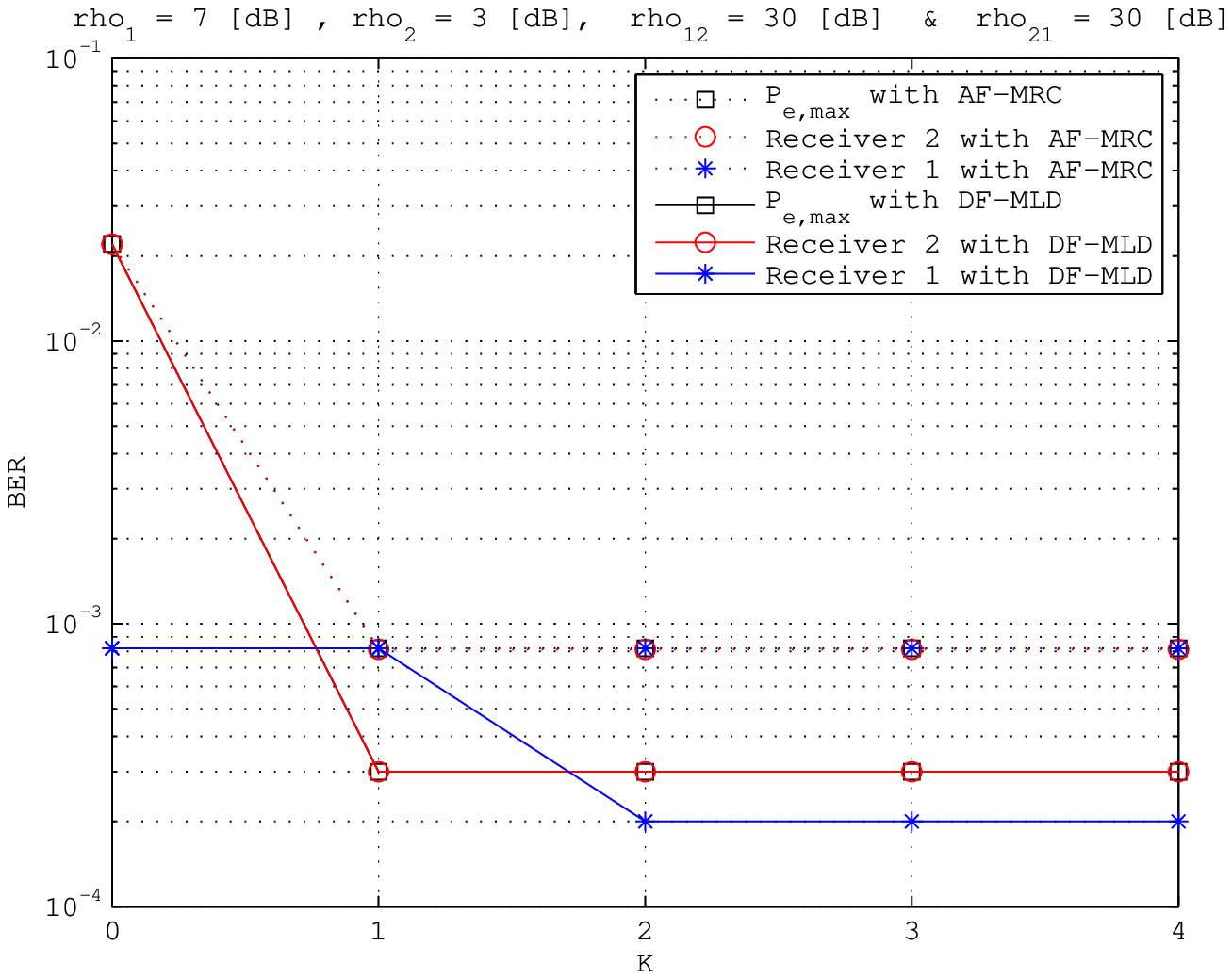}}
  \end{center}
  \caption{Performance for the asymmetric cooperation vs number of cooperation exchanges in low cooperative regime  \ (a)  \ and high cooperative regime \ (b).}
\end{figure}

Figures \ref{asym_low} and \ref{asym_high} show the performance for
receiver 1 and 2, the system performance in the low and high
cooperation regimes respectively as defined previously. Although the
system bandwidth increases with $K$, we see that the system
performance is maximum (low cooperation regime) or reaches a floor
(high cooperation regime) for two cooperation exchanges. There are
at least three reasons for this. First of all, the gain provided by
an additional cooperation round decreases with $K$. Second, the
cooperation power per exchange also decreases with $K$. In addition,
in order to derive the MLD, we have made the simplifying assumption
that the decoding errors and receive noise at each receiver are
independent, which is perfectly true for $K\leq 2$. In Figure
\ref{asym_high} we also observe the impact of the derived MLD on
system performance. We compare the DF protocol associated to MLD with the AF protocol associated to the MRC in terms of the system and individual receivers performance, and we observe that the use of the MRC limits the expected
performance gain since this combiner does not take into account the
eventual decoding made at the receivers unlike the MLD. The
observations are similar to those in \cite{djeumou-isspit-2006}.

\subsection{Comparison between the AF protocol (strategy $\mc{S}_1$) and
the DF protocol.}

\begin{figure}[h!]
  \begin{center}
    \includegraphics[height=0.40\textwidth, width=0.49\textwidth]
    {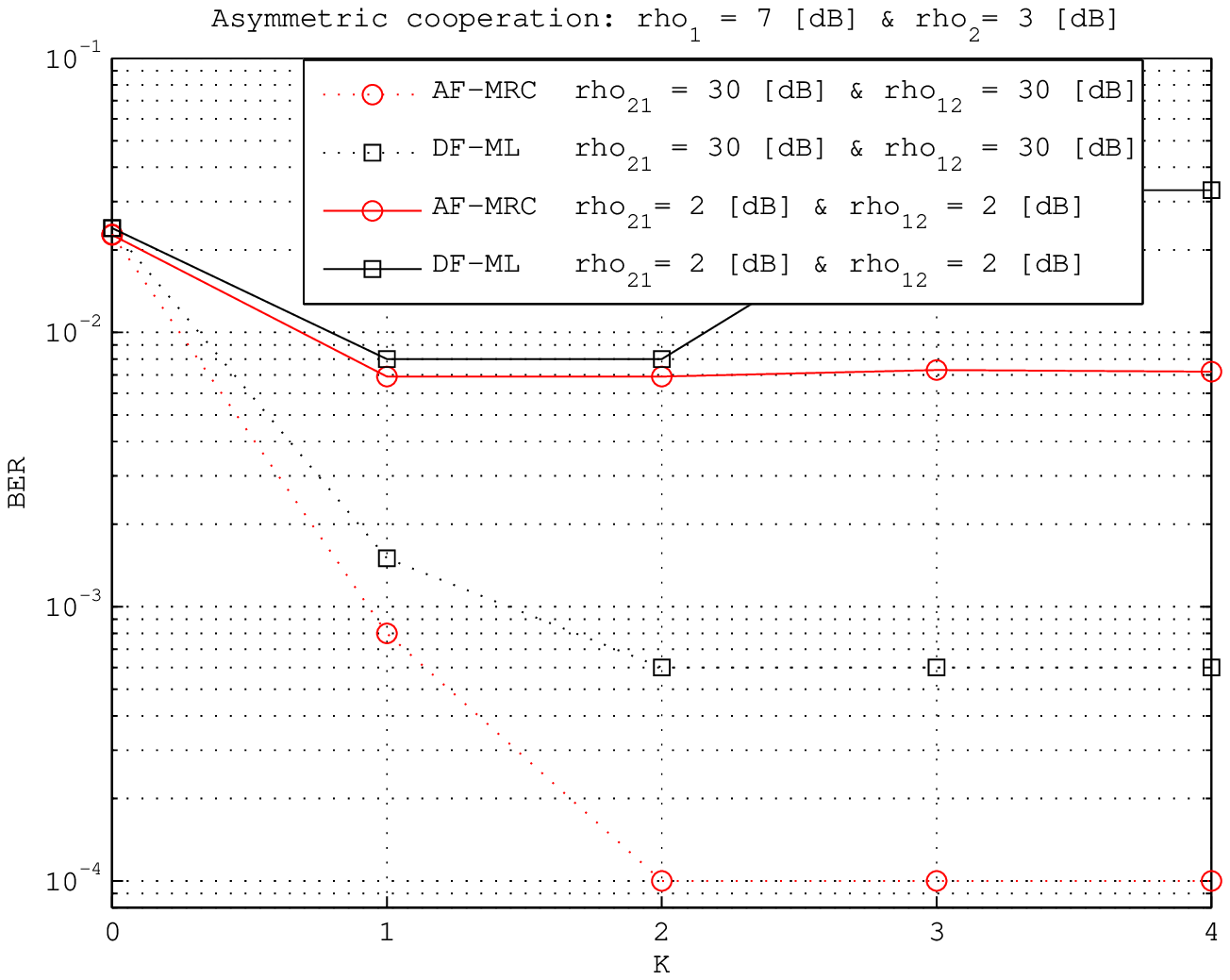}
  \end{center}
  \caption{maximum BER \ (max$\{BER_1, BER_2\}$) \ at the receivers for the AF protocol ($\mc{S}_1$) with MRC and DF protocol with MLD}
  \label{fig:af-vs-df}
\end{figure}

We consider the case where only the downlink bandwidth is fixed. We
look at the following SNR scenario: $(\frac{P}{n_1 B}, \frac{P}{n_2
B}, \frac{P_{12}}{n_{12} B}, \frac{P_{21}}{n_{21} B}, ) = (7 \
\mathrm{dB}, 3 \ \mathrm{dB}, 30 \ \mathrm{dB}, 30 \ \mathrm{dB})$
and $(\frac{P}{n_1 B}, \frac{P}{n_2 B}, \frac{P_{12}}{n_{12} B},
\frac{P_{21}}{n_{21} B}, ) = (7 \ \mathrm{dB}, 3 \ \mathrm{dB}, 2 \
\mathrm{dB}, 2 \ \mathrm{dB})$. Figure \ref{fig:af-vs-df} represent
the BER performance obtained with the AF protocol (strategy
$\mc{S}_1$) associated with the MRC and the DF protocol associated
with the MLD. We observe the impact of the hard decision with the DF
protocol which results in a performance loss in comparison to the AF
protocol. This is due to the fact that the receiver acting as the
sender does not decode perfectly the message. If one receiver can
succeed to decode the message with only the downlink signal, the DF
protocol would perform better than the AF one (see
\cite{djeumou-isspit-2006} for the same analysis on the relay
channel) and the optimal number of cooperative exchanges will
obviously be equal to $K^*=1$.

\section{Conclusions}
\label{sec:concl}


In this paper we treated four main issues inherent to the
bidirectional CBC with a single common message when power and
spectral resources are taken into account, which cannot be
considered through the discrete approach
\cite{draper-allerton-2003,khalili-allerton-2006,dabora-it-2006}.
This study was made for a simplified scenario where scalar relaying
protocols are assumed and channel coding/decoding are not exploited
(note however that one of the main practical advantages of this
approach is that the extra decoding delay induced by cooperation is
relatively small). Although we have made these simplifying
assumptions, our approach still captures the main implementation
issues posed by the bidirectional cooperation. The observations made
in this paper could be refined and used to introduce cooperation in
systems like the DVB or DVB-H systems. Here are a few key
observations we have made.

Concerning the way of \emph{combining} the signals at the receiver
we have seen that the MRC is the optimum combiner whatever the
number of cooperation rounds when the AF protocol is used. For the
DF protocol we have not only seen that an ML detector is useful
since it can compensate for the decoding noise introduced by the
other receiver but also that it is necessary to combine signals with
different constellations, which is likely to happen in practice if
the downlink and cooperation channels have different bandwidths.
Additionally the choice between sending the downlink signal or the
combiner output as a cooperation signal does not seem to be critical
for the AF protocol but the second solution complicates the
derivation of the ML detector.

\emph{Number of cooperation rounds.} By assuming the system total
bandwidth and then the downlink bandwidth to be constant, we have
seen that the system performance does not increase for more than two
cooperation rounds ($K^*\in\{0,1,2\}$), in contrast with
\cite{draper-allerton-2003,khalili-allerton-2006} for discrete
channels. We have shown for the AF protocol that the equivalent SNR
is strictly constant for $K\geq 2$ for the strategy $\mc{S}_2$ and
is almost constant or reaches its maximum for $K=1,2$ or marginally
for $3$ with the strategy $\mc{S}_1$.

\emph{Asymmetric/symmetric cooperation:} when the system bandwidth
is fixed the asymmetric cooperation has the advantage to contain the
case $K=1$ for which the best performance is generally achieved.
Indeed as the bandwidth decreases linearly with $K$ but only
logarithmically with the SNR, higher values of $K$ generally lead to
suboptimum performance. This is the main reason why the asymmetric
cooperation is preferable to the symmetric cooperation. When the
downlink bandwith is fixed, the best performance can be achieved for
$K=2$ typically. In this case, the asymmetric cooperation suffers of
a correlation effect which reduces the cooperation gain w.r.t. the
symmetric case. Additionally for $K \geq 2$ the user who starts
sending the cooperation signal has to be selected. The influence of
the available cooperation powers and noise levels at the receiver on
the best order was assessed and shown to be not negligible. In
fading channels, this order should therefore be chosen adaptively,
which is a further drawback of the asymmetric cooperation if
$K^*\geq 2$. On the other hand,  if most of the performance gain
could be captured by one cooperation round ($K^*=1$), the asymmetric
case is the best choice.

\section{Appendix}

\subsection{Conservation of the MI for the MRC}
\label{sec:proof1}
\emph{Proof:} We want to prove that $I(X;Y_{ II}^{(i)}) = I(X;
Y_{II}^{(i-1)}, Y_{12}^{(i)})$. Using the same notations as in
Section III.A. and B. we obtain that:
\begin{equation}
I(X;Y_{II}^{(i)})  = \log \left[ 1 + \frac{  \left| w_{12}^{(i)}
a_{12}^{(i)}\alpha_{I}^{(i-1)} + w_{2}^{(i)}\alpha_{II}^{(i-1)}
\right|^2 P  }{\left|w_{12}^{(i)} \right|^2 \left(
(a_{12}^{(i)})^2N_I^{(i-1)} +N_{12}^{(i)} \right) + \left|
w_2^{(i)}\right|^2 N_{II}^{(i-1)}+ w_2^{(i)}
w_{12}^{(i)}a_{12}^{(i)} (e^{(i-1)}+e^{(i-1),*})                 }
\right]
\end{equation}
Now, by replacing the MRC weights given in Proposition 1, we further
have:
\begin{equation}
I(X;Y_{II}^{(i)}) = \log \left[1 + \frac{  S_2^{(i)} }{  T_2^{(i)} } \right]
\end{equation}
where $S_2^{(i)}$ and $T_2^{(i)}$ are given by the following expressions:
\begin{equation}
S_2^{(i)} = \left( (\alpha_{II}^{(i-1)})^2   \left( (a_{12}^{(i)})^2N_I^{(i-1)} +N_{12}^{(i)} \right) + \alpha_{I}^{(i-1)}
a_{12}^{(i)}N_{II}^{(i-1)}-(a_{12}^{(i-1)})^2 \alpha_I^{(i-1)} \alpha_{II}^{(i-1)} (e^{(i-1)}+e^{(i-1),*})\right)P
\end{equation}
\begin{equation}
T_2^{(i)} = N_{II}^{(i-1)} \left( (a_{12}^{(i)})^2N_I^{(i-1)} +N_{12}^{(i)} \right)  - a_{12}^{(i)} |e^{(i-1)}|^2
\end{equation}
On the other hand we have
\begin{equation}
\begin{array}{ccl}
I(X;Y_{II}^{(i-1)}, Y_{12}^{(i)} ) & = & I(X;
(\alpha_{II}^{(i-1)}X+Z_{II}^{(i-1)}),
(a_{12}^{(i)}\alpha_{I}^{(i-1)}X + a_{12}^{(i)}Z_I^{(i-1)}
+Z_{12}^{(i)} )) \\
 & = & \log \left[1 + \frac{  \tilde{S}_2^{(i)} }{  \tilde{T}_2^{(i)} } \right].
\end{array}
\end{equation}
We see that $\tilde{S}_2^{(i)} = S_2^{(i)}$ and $\tilde{T}_2^{(i)} = S_2^{(i)}$, which concludes the proof
\subsection{Proof of theorem \ref{theo-snr}}
\label{sec:proof2}
Here we show the result obtained for the equivalent SNR in Theorem
\ref{theo-snr}. At receiver 2, for the iteration $i \in
\{1,...,K\}$, the signal at the MRC output is denoted by
$Y_{II}^{(i)} = \alpha_{II}^{(i)} X + Z_{II}^{(i)}$. The
equivalent SNR $\rho_{II}^{(i)}$ in $Y_{II}^{(i)}$ expresses as
$\rho_{II}^{(i)} \triangleq \frac{E\left|\alpha_{II}^{(i)}
X\right|^2}{E\left|Z_{II}^{(i)}\right|^2} =
\frac{\mc{S}^{(i)}}{\mc{T}^{(i)}}$ with
\begin{eqnarray}
\mc{S}^{(i)}    &=& E\left|\alpha_{II}^{(i)} X\right|^2 \nonumber\\
                &=& \left[w_{12}^{(i)}a_{12}^{(i)}\alpha_{I}^{(i-1)} + w_2^{(i)}
\alpha_{II}^{(i-1)}\right]^2 P \nonumber\\
                &=& \left[\left(a_{12}^{(i)}\alpha_{I}^{(i-1)}N_{II}^{(i-1)}-  a_{12}^{(i)}
 \alpha_{II}^{(i-1)} e^{(i-1)}\right)a_{12}^{(i)}\alpha_{I}^{(i-1)} \right. \nonumber\\
                & & + \left. \left([\left(a_{12}^{(i)}\right)^2N_{I}^{(i-1)}+N_{12}^{(i)} ]
 \alpha_{II}^{(i-1)} - \left(a_{12}^{(i)}\right)^2 \alpha_{I}^{(i-1)} e^{(i-1)}\right)
\alpha_{II}^{(i-1)}\right]^2 P \nonumber\\
                &=& \left[\left(\alpha_{I}^{(i-1)}\alpha_{II}^{(i-1)}\right)^2PN_{12} + \left(\alpha_{I}^{(i-1)}\right)^2N_{II}^{(i-1)}P_{12} + \left(\alpha_{II}^{(i-1)}\right)^2N_{I}^{(i-1)}P_{12}  - \alpha_{I}^{(i-1)}\alpha_{II}^{(i-1)}P_{12}\left(e^{(i-1)}+e^{(i-1),*}\right) \right. \nonumber\\
                & & + \left. \left(\alpha_{II}^{(i-1)}\right)^2N_{I}^{(i-1)}N_{12} \right] \nonumber\\
                & & \cdot \underbrace{\left[\left(a_{12}^{(i)}\alpha_{I}^{(i-1)}\right)^2N_{II}^{(i-1)} + \left(a_{12}^{(i)}\alpha_{II}^{(i-1)}\right)^2N_{I}^{(i-1)} + \left(\alpha_{II}^{(i-1)}\right)^2N_{12} - \left(a_{12}^{(i)}\right)^2\alpha_{I}^{(i-1)}\alpha_{II}^{(i-1)}\left(e^{(i-1)}+e^{(i-1),*}\right)\right]}_{\mc{C}} P \nonumber
\end{eqnarray}
and,
\begin{eqnarray}
\mc{T}^{(i)}    &=& E\left|Z_{II}^{(i)}\right|^2 \nonumber\\
                &=& \left(w_{12}^{(i)}a_{12}^{(i)} \right))^2 N_{I}^{(i-1)} + \left(w_{12}^{(i)}\right)^2 N_{12}^{(i)}
+\left(w_{2}^{(i)}\right)^2N_{II}^{(i-1)}  +2 w_{12}^{(i)} w_{2}^{(i)}\left(e^{(i-1)}+e^{(i-1),*}\right) \nonumber \\
                &=&  \left(a_{12}^{(i)}\alpha_{I}^{(i-1)}N_{II}^{(i-1)}-  a_{12}^{(i)}
 \alpha_{II}^{(i-1)} e^{(i-1)}\right)^2 \left(\left(a_{12}^{(i)}\right)^2N_{I}^{(i-1)} +   N_{12}^{(i)}\right) \nonumber\\
                & & +\left(\left[(a_{12}^{(i)})^2N_{I}^{(i-1)}+N_{12}^{(i)} \right]
 \alpha_{II}^{(i-1)} - (a_{12}^{(i)})^2 \alpha_{I}^{(i-1)} e^{(i-1)}\right)^2N_{II}^{(i-1)}  \nonumber\\
                & & + 2  \left([(a_{12}^{(i)})^2N_{I}^{(i-1)}+N_{12}^{(i)}]
 \alpha_{II}^{(i-1)} - (a_{12}^{(i)})^2 \alpha_{I}^{(i-1)} e^{(i-1)}\right) \nonumber \\
                                & & \hspace{0.4cm}\cdot\left(a_{12}^{(i)}\alpha_{I}^{(i-1)}N_{II}^{(i-1)}-  a_{12}^{(i)}
 \alpha_{II}^{(i-1)} e^{(i-1)}\right)\left(e^{(i-1)}+e^{(i-1),*}\right) \nonumber \\
                &=& \left[P_{12}e^{(i-1)}e^{(i-1),*} - \left(\alpha_{I}^{(i-1)}\right)^2PN_{II}^{(i-1)}N_{12} - N_{I}^{(i-1)}N_{II}^{(i-1)}\left(P_{12} + N_{12}\right)\right] \nonumber\\
                & & \cdot \underbrace{\left[\left(a_{12}^{(i)}\alpha_{I}^{(i-1)}\right)^2N_{II}^{(i-1)} + \left(a_{12}^{(i)}\alpha_{II}^{(i-1)}\right)^2N_{I}^{(i-1)} + \left(\alpha_{II}^{(i-1)}\right)^2N_{12} - \left(a_{12}^{(i)}\right)^2\alpha_{I}^{(i-1)}\alpha_{II}^{(i-1)}\left(e^{(i-1)}+e^{(i-1),*}\right)\right]}_{\mc{C}} \nonumber
\end{eqnarray}
After simplification w.r.t. the common factor $\mc{C}$, we obtain $\rho_{II}^{(i)} = \frac{\mc{S}_{I}^{(i)}}{\mc{T}_{I}^{(i)}}$ with
$$
\begin{array}{ccl}
\mc{S}_{I}^{(i)}    &=& \left[\left(\alpha_{I}^{(i-1)}\alpha_{II}^{(i-1)}\right)^2PN_{12} + \left(\alpha_{I}^{(i-1)}\right)^2N_{II}^{(i-1)}P_{12} + \left(\alpha_{II}^{(i-1)}\right)^2N_{I}^{(i-1)}P_{12}  - \alpha_{I}^{(i-1)}\alpha_{II}^{(i-1)}P_{12}\left(e^{(i-1)}+e^{(i-1),*}\right) \right. \\
                & & \left. + \left(\alpha_{II}^{(i-1)}\right)^2N_{I}^{(i-1)}N_{12} \right]P \\ \\
\mc{T}_{I}^{(i)}    &=& \left[P_{12}e^{(i-1)}e^{(i-1),*} - \left(\alpha_{I}^{(i-1)}\right)^2PN_{II}^{(i-1)}N_{12} - N_{I}^{(i-1)}N_{II}^{(i-1)}\left(P_{12} + N_{12}\right)\right]
\end{array}
$$
Then, by multiplying both the numerator and the denominator of $\rho_{II}^{(i)}$ by the factor $\ds{ \frac{\rho_{I}^{(i-1)}\rho_{II}^{(i-1)}}{PN_{12}} }$, we obtain $\rho_{II}^{(i)}
 = \ds{ \frac{\mc{S}_{II}^{(i)}}{\mc{T}_{II}^{(i)}} }$ with
\begin{equation}
\begin{array}{ccl}
\mc{S}_{II}^{(i)}  &=& \ds{\alpha_{I}^{(i-1)}\alpha_{II}^{(i-1)}
\left(e^{(i-1)}+e^{(i-1),*}\right)\rho_{I}^{(i-1)}\rho_{II}^{(i-1)}\rho_{12}}
 \ds{-\left(\alpha_{I}^{(i-1)}\alpha_{II}^{(i-1)}\right)^2P
 \left[\rho_{II}^{(i-1)}\left(1+\rho_{I}^{(i-1)}\right) \right.} \\
 && \ds{+ \left.
 \rho_{12}\left(\rho_{I}^{(i-1)}+\rho_{II}^{(i-1)}\right)\right]}\\ \\
\mc{T}_{II}^{(i)} & = & \ds{
\frac{e^{(i-1)}e^{(i-1),*}}{P}\rho_{I}^{(i-1)}\rho_{II}^{(i-1)}\rho_{12}
} \ds{-
\left(\alpha_{I}^{(i-1)}\alpha_{II}^{(i-1)}\right)^2P\left(1 +
\rho_{12}\right)} - \ds{\left(\alpha_{I}^{(i-1)}\right)^2
 N_{II}^{(i-1)}  \rho_{I}^{(i-1)}\rho_{II}^{(i-1)}}.
\end{array}
\end{equation}\\
\subsection{Proof of Proposition \ref{prop-symm}: symmetric cooperation}
\label{sec:proof3}
Here we only show how to obtain the MRC weights, and this for receiver 2. The signal
coefficients and equivalent noises can be derived from the equivalent signal expressions.\\
At receiver 2, at the iteration $i$, the signals available at the combiner inputs are
\begin{eqnarray}
\left\{
    \begin{array}{ccll}
    Y_{12}^{(i)} & = & a_{12}^{(i)} Y_{I}^{(i-1)} + Z_{12}^{(i)}  = a_{12}^{(i)} \alpha_{I}^{(i-1)} X + (a_{12}^{(i)} Z_{I}^{(i-1)} + Z_{12}^{(i)})  \\
    Y_{II}^{(i-1)} & = & \alpha_{II}^{(i-1)} X + Z_{II}^{(i-1)}  \\
    \end{array}
\right. \nonumber
\end{eqnarray}

Denote by $\bs{w}^{(i)} = (w_{12}^{(i)}, w_{2}^{(i)})^{t}$ the
optimal weight vector. For the maximum ratio combiner,
$\bs{w}^{(i)}$ is given by the following expression:
\begin{eqnarray}
    \bs{w}^{(i)}  =  {\bs{R}_{zz}^{(i)}}^{-1} \cdot \bs{h}^{*}
\end{eqnarray}
where $\bs{R}_{zz}^{(i)}$ is the covariance matrix between the equivalent noises $(a_{12}^{(i)} Z_{I}^{(i-1)} + Z_{12}^{(i)})$ and $Z_{II}^{(i-1)}$ with
\begin{equation}
\left\{
\begin{array}{ccc}
Z_{I}^{(i-1)} &=& w_{21}^{(i-1)}a_{21}^{(i-1)}Z_{II}^{(i-1)} +
w_{21}^{(i-1)}Z_{21}^{(i-1)}+w_{1}^{(i-1)}Z_{I}^{(i-2)} \\
Z_{II}^{(i-1)} &=& w_{12}^{(i-1)}a_{12}^{(i-1)}Z_{I}^{(i-2)} +
w_{12}^{(i-1)}Z_{12}^{(i-1)}+w_{2}^{(i-1)}Z_{II}^{(i-2)}.
\end{array}
 \right. \nonumber
\end{equation}
and $\bs{h}$ is the useful signal coefficients vector given by $\bs{h} = (a_{12}^{(i)} \alpha_{I}^{(i-1)}, \alpha_{II}^{(i-1)})^{t}$ with $\alpha_{I}^{(i-1)}$ and $\alpha_{II}^{(i-1)}$ obtained from the following recursive formula
\begin{equation}
\left\{
\begin{array}{ccl}
\alpha_{I}^{(i)} &= &w_{21}^{(i)}a_{21}^{(i)}\alpha_{II}^{(i-1)} +
w_{1}^{(i)} \alpha_I^{(i-1)} \\
\alpha_{II}^{(i)} &= &w_{12}^{(i)}a_{12}^{(i)}\alpha_{I}^{(i-1)} +
w_{2}^{(i)} \alpha_{II}^{(i-1)}.
\end{array}
\right. \end{equation}

Therefore we find that
\begin{eqnarray}
\bs{R}_{zz}^{(i)} = \left[
    \begin{array}{ccc}
    \left(a_{12}^{(i)}\right)^{2}N_{I}^{(i-1)} + N_{12}^{(i)} &  & a_{12}^{(i)} e^{(i-1)} \\
    a_{12}^{(i)} e^{(i-1),*} &  &  N_{II}^{(i-1)}
    \end{array}
    \right] \nonumber
\end{eqnarray}

and

\begin{eqnarray}
{\bs{R}_{zz}^{(i)}}^{-1} = \frac{1}{\left[\left(a_{12}^{(i)}\right)^{2}N_{I}^{(i-1)} + N_{12}^{(i)}\right]N_{II}^{(i-1)} - |e^{(i-1)}|^{2}} \left[
    \begin{array}{ccc}
    N_{II}^{(i-1)} &   & - a_{12}^{(i)} e^{(i-1)} \\
    - a_{12}^{(i)} e^{(i-1),*} &   & \left(a_{12}^{(i)}\right)^{2}N_{I}^{(i-1)} + N_{12}^{(i)}
    \end{array}
    \right] \nonumber
\end{eqnarray}
where $e^{(i-1)} \triangleq E \left[ Z_{I}^{(i-1)} Z_{II}^{(i-1),*}  \right]$.\\
Finally we obtain
\begin{equation}
\label{eq:weights1} \left\{
\begin{array}{ccl}
 w_{12}^{(i)} &= &  a_{12}^{(i)}\alpha_{I}^{(i-1)}N_{II}^{(i-1)}-  a_{12}^{(i)}
 \alpha_{II}^{(i-1)} e^{(i-1)}
 \\
 w_{2}^{(i)} & = & \left[(a_{12}^{(i)})^2N_{I}^{(i-1)}+N_{12}^{(i)} \right]
 \alpha_{II}^{(i-1)} - (a_{12}^{(i)})^2 \alpha_{I}^{(i-1)} e^{(i-1),*} \\
\end{array}
\right.
\end{equation}

and we start with  $e^{(0)} = 0$,
$N_{I}^{(0)} = N_1$, $N_{II}^{(0)} = N_2$, $\rho_{I}^{(0)}
=P/N_1$, $\rho_{II}^{(0)} =P/N_2$, $\alpha_{I}^{(0)} =
\alpha_{II}^{(0)} = 1 $.
\subsection{Proof of Proposition \ref{prop-asymm}}
\label{sec:proof4}
Compared to the symmetric case, only the
 equivalent noise expressions and the useful signal coefficients are changed. They
 can be
 obtained from the signal expressions (\ref{eq:received-af})  and
 shown to be
\begin{equation}
\alpha_{I}^{(i)} = \left|
\begin{array}{cc}
w_{21}^{(i)}a_{21}^{(i)}\alpha_{II}^{(i-1)}+w_{1}^{(i)}
\alpha_{I}^{(i-1)} & \ for \ i \ even \\
\alpha_{I}^{(i-1)} & \ for \ i \ odd,
\end{array}
 \right.
\end{equation}
\begin{equation} \alpha_{II}^{(i)} = \left|
\begin{array}{cc}
\alpha_{II}^{(i-1)} & \ for \ i \  even \\
w_{12}^{(i)}a_{12}^{(i)}\alpha_{I}^{(i-1)}+w_{2}^{(i)}\alpha_{II}^{(i-1)}
& \ for \ i \ odd,
\end{array}
\right.
\end{equation}
\begin{equation}
Z_{I}^{(i)} = \left|
\begin{array}{cc}
w_{21}^{(i)}a_{21}^{(i)}Z_{II}^{(i-1)} +
w_{21}^{(i)}Z_{21}^{(i)}+w_{1}^{(i)}Z_{I}^{(i-1)}
&  \ i \ even \\
Z_{I}^{(i-1)} & \ i \  odd,
\end{array}
 \right. \nonumber
\end{equation}

\begin{equation}
Z_{II}^{(i)} = \left|
\begin{array}{cc}
Z_{II}^{(i-1)}
&  \ i \ even \\
w_{12}^{(i)}a_{12}^{(i)}Z_{I}^{(i-1)} +
w_{12}^{(i)}Z_{12}^{(i)}+w_{2}^{(i)}Z_{II}^{(i-1)}
 & \ i \  odd,
\end{array}
 \right. \nonumber
\end{equation}
Doing the same calculation as for the previous proposition leads
to the MRC weights which have the same expressions as in the
symmetric case.

%
%
%
%
%
%

\end{document}